\documentclass[a4paper,11pt]{article}
\pdfoutput=1 

\usepackage{jheppub} 

\usepackage[T1]{fontenc} 

\usepackage{amsmath}    
\usepackage{amssymb}    %
\usepackage{graphicx}   
\usepackage{verbatim}   
\usepackage{color}      
\usepackage{subfigure}  
\usepackage{hyperref}
\usepackage{multirow}
\usepackage{comment}
\usepackage{pdflscape}
\usepackage{rotating}
\usepackage{float}	
\usepackage{color}
\definecolor{rosso}{RGB}{210,0,0}

\newcommand\sss{\scriptscriptstyle}

\newcommand{\gev}{\,\textrm{GeV}}

\newcommand{\TO}{\rightarrow}

\newcommand{\tth}{t\bar{t}H}

\newcommand{\asa}[2]{\alpha_s^{#1}\alpha^{#2}}
\newcommand{\ord}{\mathcal{O}}
\newcommand{\tril}{\lambda_{3}}
\newcommand{\trilsm}{\tril^{\rm SM}}
\newcommand{\qual}{\lambda_{4}}
\newcommand{\mh}{m_{ \sss H}}
\newcommand{\mw}{m_{ \sss W}}
\newcommand{\mz}{m_{ \sss Z}}
\newcommand{\mt}{m_{t}}
\newcommand{\hw}{h_{\sss W}}

\newcommand{\ktre}{\kappa_{\lambda}}
\newcommand{\kqual}{\kappa_{\lambda_4}}
\newcommand{\dsigma}{\delta \Sigma_{\tril}}
\newcommand{\dsigmah}{\delta \sigma_{\tril}}
\newcommand{\dgamma}{\delta \Gamma_{\tril}}
\newcommand{\dBR}{\delta {\rm BR}_{\tril}}
\newcommand{\ggF}{gg{\rm F}}
\def\beq{\begin{equation}}
\def\beqn{\begin{eqnarray}}
\def\eeq{\end{equation}}
\def\eeqn{\end{eqnarray}}
\def\beal{\begin{align}}
\def\endal{\end{align}}

\newcommand{\br}{{\rm BR}}
\newcommand{\muif}{\mu_i^f}
\newcommand{\muifexp}{\bar{\mu}_i^f}

\newenvironment{appendletterA}
 {
  \typeout{ Starting Appendix \thesection }
  \setcounter{section}{0}
  \setcounter{equation}{0}
  
 }{
  \typeout{Appendix done}
 }

\newenvironment{appendletterB}
 {
  \typeout{ Starting Appendix \thesection }
  \setcounter{equation}{0}
  
 }{
  \typeout{Appendix done}
 }

\vspace*{-1cm}
\begin{document}\hfill
{\flushright{
        \begin{minipage}{3.0cm}
          CP3-16-38 \\
          RM3-TH/16-8
        \end{minipage}        }

}
\vspace*{1cm}
\color{black}
\begin{center}
{\Large \bf  \color{blue} Probing the Higgs self coupling\\[8pt] via single Higgs production at the LHC}

\bigskip\color{black}\vspace{.4cm}
{\large\bf G.~Degrassi$^a$, P.P.~Giardino$^{b}$, F.~Maltoni$^c$, D.~Pagani$^c$}
\\[7mm]
{\it  (a) Dipartimento di Matematica e Fisica, Universit{\`a} di Roma Tre and \\
 INFN, sezione di Roma Tre, I-00146 Rome, Italy}\\[1mm]
 {\it (b) Physics Department, Brookhaven National Laboratory, \\Upton, New York
11973, US}\\[1mm]
{\it (c)  Centre for Cosmology,  Particle Physics and Phenomenology (CP3),
Universit\'e Catholique de Louvain, B-1348 Louvain-la-Neuve, 
Belgium}\\[1mm]
\end{center}
\bigskip
\vspace{.5cm}

\centerline{\large\bf Abstract}
\begin{quote}
We propose a method to determine the trilinear Higgs self coupling that is alternative to the direct measurement of Higgs pair production total cross sections and differential distributions. The method relies on the effects that electroweak loops featuring an anomalous trilinear coupling would imprint on single Higgs production at the LHC. We first calculate these contributions to all the phenomenologically relevant Higgs production ($\ggF$, VBF, $WH$, $ZH$, $t\bar tH$) and decay ($\gamma \gamma$, $WW^{*}/ZZ^{*}\to 4f$, $b\bar b$, $\tau \tau$) modes at the LHC and then estimate the sensitivity to the trilinear coupling via a one-parameter fit to the single Higgs measurements at the LHC 8 TeV.  We find that the bounds on  the self coupling are already competitive with those from Higgs pair production and will be further improved in the current and next LHC runs.
\end{quote}
\thispagestyle{empty}
\newpage

\section{Introduction\label{sec:intro}} 

The discovery of a new scalar resonance with mass around 125 GeV at
the Large Hadron Collider (LHC)\cite{Chatrchyan:2012xdj, Aad:2012tfa}
opened a new era in high-energy particle physics. The study of the
properties of this particle provides strong evidence that it is the
Higgs boson of the Standard Model (SM), {\it i.e.}, a scalar CP-even
state whose couplings to the other known particles have a SM-like
structure and strengths proportional to their masses.  In particular,
ATLAS and CMS performed both independent \cite{Khachatryan:2014jba,
  Aad:2015gba} and combined \cite{Khachatryan:2016vau} studies on the
Higgs couplings in the so-called
$\kappa$-framework~\cite{LHCHiggsCrossSectionWorkingGroup:2012nn,Heinemeyer:2013tqa},
where the predicted SM Higgs strengths $c_{i}$ are rescaled by overall
factors $\kappa_i$. In the combined analysis based on 7 and 8 TeV data
sets \cite{Khachatryan:2016vau} the couplings with the vector bosons
have been found to be compatible with those expected from the SM, {\it
  i.e.}, $\kappa_V =1$ $(V=W,Z)$, within a $\sim10\%$ uncertainty,
while in the case of the heaviest SM fermions (the top, the bottom
quarks and the $\tau$ lepton) the uncertainty is of order
$\sim15-20\%$. However, at this stage, additional relations among the
different $\kappa_i$ that improve the sensitivity of experimental
analyses are often assumed, yet lead to a loss of generality.  The
precision of the current measurements therefore still leaves room for
Beyond-the-Standard-Model (BSM) scenarios involving modifications of
the Higgs boson couplings to the vector bosons and fermions.

Besides the direct search of new particles, one of the main tasks of
the second run of the LHC at $\sqrt{s} = 13$ TeV centre-of-mass energy
will be the precise determination of the properties and the
interactions of the SM particles, in particular those of the Higgs
boson, in order to constrain effects from New Physics (NP).  The
increase of the production cross sections together with a larger
integrated luminosity, which is expected to reach 300 fb$^{-1}$ per
experiment at the end of the Run II and up to 3000 fb$^{-1}$ in the
case of the following High Luminosity (HL) option, will allow to probe
the couplings of the Higgs boson with the other SM particles with much
higher accuracy. In particular, present estimates \cite{CMS:2013xfa,
  Peskin:2013xra}, suggest that at the end of Run II the Higgs boson
couplings to the vector bosons are expected to reach a $\sim 5 \%$
precision with 300 fb$^{-1}$ luminosity, while the couplings to the
heavy fermions could reach $\sim 10-15 \%$ precision. Similar
estimates for the end of the HL option indicate a reduction of these
numbers by at least a factor $\sim 2$.


The study of the trilinear $(\tril)$ and quartic $(\qual)$ Higgs self
couplings in the scalar potential 
\begin{eqnarray}
V(H) =  \frac{\mh^2}{2} H^2 + \lambda_3 v  H^3 + \lambda_4 H^4  \nonumber
\end{eqnarray}
is in a completely different situation. In the SM, the potential is
fully determined by only two parameters,  $v=(\sqrt2
G_\mu)^{-1/2}$ and the coefficient of the $(\Phi^\dagger \Phi)^2$
interaction $\lambda$, where $\Phi$ is the Higgs doublet field. Thus,
the mass and the self couplings of the Higgs boson depend only on
$\lambda$ and $v$ ($\mh^2=2 \lambda v^2, \trilsm = \lambda, \qual^{\rm
  SM} = \lambda/4$). On the contrary, in the case of extended scalar
sectors or in presence of new dynamics at higher scales the trilinear
and quartic couplings, $\tril$ and $\qual$, typically depend on
additional parameters and their values can depart from the SM
predictions~\cite{Gupta:2013zza,Efrati:2014uta}.

At the Leading Order (LO) the Higgs decay widths and the cross sections of the main single Higgs production processes, {\it i.e.}, gluon--gluon fusion ($\ggF$), vector-boson fusion (VBF), $W$ and $Z$ associated production ($WH$, $ZH$) and the production in association with a top-quark pair ($t\bar tH$),  depend on the couplings of the Higgs boson to the other particles of the SM, yet they are insensitive to $\tril$ and
$\qual$.  Information on $\tril$ can be directly obtained at LO only from final states featuring at least two Higgs bosons.  However, the cross sections of these processes are
much smaller than those of single Higgs production, due to the
suppression induced by a heavier final state and an additional weak
coupling. At $\sqrt{s} = 13 $ TeV the single Higgs
gluon-gluon-fusion production cross section in the SM is around 50 pb
\cite{Anastasiou:2016cez}, while the double Higgs cross section is
around 35 fb in the gluon-gluon-fusion channel~\cite{deFlorian:2013jea,Maltoni:2014eza,Borowka:2016ehy} and even smaller in other production mechanisms~\cite{Baglio:2012np,Frederix:2014hta}.

A plethora of perspective studies performed at $\sqrt{s}=13$ TeV
suggest that it should be possible to detect the production of a Higgs
pair via $b\bar{b} \gamma \gamma$~\cite{Baur:2003gp, Baglio:2012np,
  Yao:2013ika, Barger:2013jfa, Azatov:2015oxa, Lu:2015jza}, $b\bar{b}
\tau \tau$~\cite{Dolan:2012rv, Baglio:2012np}, $b\bar{b}W^+
W^-$~\cite{Papaefstathiou:2012qe} and
$b\bar{b}b\bar{b}$~\cite{deLima:2014dta, Wardrope:2014kya,
  Behr:2015oqq} final states, and also via signatures emerging from $t
\bar t HH$ \cite{Englert:2014uqa, Liu:2014rva} and $HVV$
\cite{Cao:2015oxx} production channels. However, the ultimate
precision that could be achieved on the determination of $\tril$ is
much less clear. Even with an integrated luminosity of 3000 fb$^{-1}$,
experimental analyses suggest that it will be possible to exclude at
the LHC only values in the range $\tril<-1.3 ~\tril^{\rm SM}$ and
$\tril>8.7 ~ \tril^{\rm SM}$ via the $b\bar{b} \gamma \gamma$
signatures \cite{ATL-PHYS-PUB-2014-019} or $\tril<-4 ~\tril^{\rm SM}$
and $\tril>12 ~ \tril^{\rm SM}$ even including also $b\bar{b} \tau
\tau$ signatures \cite{ATL-PHYS-PUB-2015-046}, {\it i.e.}, a much more
pessimistic perspective than the results reported in the
phenomenological explorations. The current experimental bounds on
non-resonant Higgs pair production cross sections as obtained at 8 TeV
are rather weak. ATLAS has been able to exclude only a signal up to 70
times larger than the SM one \cite{Aad:2015xja,Aad:2015uka}, which can
be roughly translated to the $\tril<-12 ~\tril^{\rm SM}$ and $\tril >
17 ~ \tril^{\rm SM}$ exclusion limits, while CMS puts a 95$\%$ C.L.~
exclusion limit on $\tril< -17.5 ~\tril^{\rm SM}$ and $\tril > 22.5 ~
\tril^{\rm SM}$ assuming changes only in the trilinear Higgs boson
coupling, with all other parameters fixed to their SM values
\cite{Khachatryan:2016sey}. Thus, additional strategies in the
determination of the trilinear Higgs self coupling $\tril$ that are
alternative and complementary to the constrains from Higgs pair
production would be certainly helpful. Finally, the perspectives of
determining the quartic Higgs self coupling $\qual$ via measurements
in triple Higgs production seems quite bleak at the
LHC\cite{Plehn:2005nk, Binoth:2006ym}, due to the smallness of the
corresponding cross section~\cite{Maltoni:2014eza}.

In this work we explore the possibility of constraining the trilinear
Higgs self coupling with a different approach, namely, via precise
measurements of processes featuring single Higgs production and decay
at the LHC. Indeed, although single Higgs production does not depend
on $\tril$ at LO or at higher orders in QCD, it does depend on $\tril$
via weak loops, namely at Next-to-Leading (NLO) in the electroweak
(EW) interactions.  We therefore extract the $\tril$-dependent part
from the NLO EW corrections to all phenomenologically relevant single
Higgs production cross sections ($\ggF$, VBF, $WH$, $ZH$, $t\bar t H$)
and branching ratios, ($H\TO\gamma\gamma$, $H\TO ZZ^*,WW^*\TO 4f$,
$H\TO f\bar{f}$, $H\TO gg$).  By varying the value of $\tril$, we
evaluate the impact of an anomalous trilinear Higgs self coupling on
the predictions for the aforementioned cross sections and decay
widths. We obtain a distinctive pattern of deformations of the SM
predictions for the rates ($\sigma(i) \cdot {\rm BR}(f)$), which can
be compared to the experimental data. A similar investigation,
specific to $ZH$ production at an $e^+ e^-$ collider, was presented in
Ref.\cite{McCullough:2013rea}.
  
Our approach builds on the assumption that NP couples to the SM via
the Higgs potential and dominantly affects only the Higgs self
couplings. In other words, the lowest-order Higgs couplings to the
other fields of the SM (and in particular to the top quark and vector
bosons) are still given by the SM prescriptions or, equivalently,
modifications to these couplings are so small that do not swamp the
NLO effects we are considering. While this assumption needs always to
be kept in mind, we stress that {\it all} the current experimental
limits or estimates of limits on $\tril$ obtained from Higgs pair
production implicitly rely on it, too. In particular, the
top-quark-Higgs coupling is assumed to be the one of the
SM. Perspectives on measurements of $\tril$ via Higgs pair production
relaxing this assumption have been studied at the phenomenological
level, {\it e.g.}, in Refs.~\cite{Goertz:2014qta, Azatov:2015oxa}
leading, in general, to much weaker bounds.  Within the assumption
that NP modifies only $\tril$, we investigate the reach of our
approach in the determination of $\tril$ by considering the current 8
TeV Higgs data~\cite{Khachatryan:2016vau} and the expected
performances of the forthcoming runs of the LHC~\cite{CMS:2013xfa,
  Peskin:2013xra}.  We demonstrate the potential of single Higgs
production channels in setting bounds on $\tril$ that are competitive
and complementary to those achievable via the searches for double
Higgs production.

The paper is organised as follows. In Section \ref{sec:org} we present the theoretical 
framework and discuss the $\tril$-dependent part of the NLO EW corrections to 
the single Higgs processes. In the following section we
present the calculation of such contributions  to the various 
observables. Section \ref{sec:results} is devoted to study the impact of the
$\tril$-dependent contribution in the single Higgs production and decay modes
at the LHC, while in the following section we discuss the constraints on $\lambda_3$ that can be obtained from the current data and also from future measurements. In the last section we summarise and draw our conclusions. 

\section{$\tril$-dependent contributions in single Higgs processes}
\label{sec:org}

As basic assumption, we consider a BSM scenario where 
the only (or dominant) modification of the SM Lagrangian at low energy appears in the scalar potential. 
In other words, we  assume that the only relevant effect induced at the weak scale by unknown NP at a high scale is a modification of the self couplings of the 125 GeV boson.
In particular,  we concentrate on the trilinear
self-coupling of the Higgs boson, making the  
assumption that modifications of $\qual$ and of possible other 
self-couplings in the potential lead to much smaller effects and that 
the strength of tree-level interactions of the Higgs field  with the 
vector  bosons and with the fermions is not (or very weakly) modified with respect to the SM case.
We therefore simply parametrise the effect of NP at the weak scale via a single parameter $\ktre$, {\it i.e.},  
the rescaling of the SM trilinear coupling, $\trilsm$.  Thereby, the $H^3$ interaction in the 
potential, where $H$ is the physical Higgs field, is given by
\beq
V_{H^{3}} = \tril\, v \,H^3 \equiv \ktre \trilsm \, v \, H^3, \qquad
\trilsm = \frac{G_\mu}{\sqrt{2}} \mh^2 \,,
\label{h3coeff}
\eeq
with the vacuum expectation value, $v$, related to the Fermi constant at
the tree-level by $v= (\sqrt{2}\, G_\mu)^{-1/2}$.  

As we will discuss and quantify in more detail in the following, the ``deformation" of the Higgs trilinear coupling  induces  modifications of the Higgs couplings to fermions and to vector bosons at one loop.
However, since such loop-induced $\tril$-dependent contributions are energy- and observable-dependent, the resulting modifications cannot be parameterised via a rescaling of the tree-level couplings of the single Higgs production and decay processes considered. Thus, it is important to keep in mind that the effects discussed in this work cannot be correctly captured by the standard $\kappa$-framework~\cite{LHCHiggsCrossSectionWorkingGroup:2012nn,Heinemeyer:2013tqa}. 

\begin{figure}[t]
\begin{center}\vspace*{-1.0cm}
\includegraphics[width=0.49\textwidth]{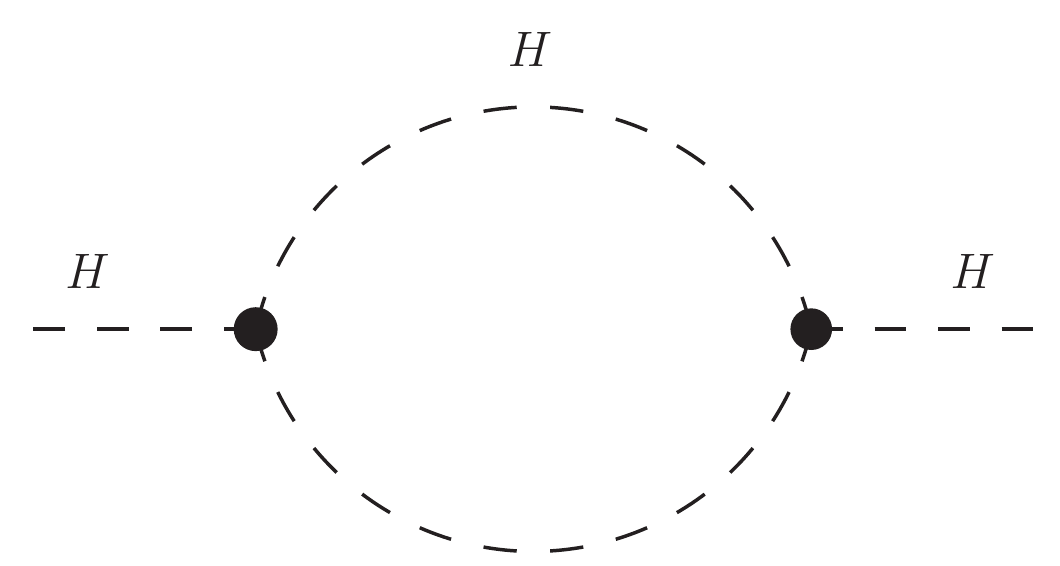}
\caption{One-loop $\tril$-dependent diagram in  the Higgs self-energy.}
\label{fig1}
\end{center}
\end{figure}

Let us now start by classifying the $\tril$-dependent contributions that come from the
$\ord(\alpha)$ corrections  to single Higgs production and decay processes. 
These contributions can be divided into two categories: a universal part, {\it i.e.}, common to all processes,  quadratically dependent on $\tril$ and a process-dependent part linearly proportional to $\tril$.

The universal $\ord({\tril}^2)$ corrections originate from the diagram in the wave function renormalisation constant of the external Higgs field, see Fig.~\ref{fig1}.  This contribution represents a  renormalisation factor common to all the vertices where the Higgs couples to vector bosons or fermions. Thus, for on-shell Higgs boson production and decay, it induces the same effect for all processes, without any dependence on the kinematics.   Denoting as $\mathcal{M}$ a generic amplitude for single Higgs production or a Higgs decay width, the correction to $\mathcal{M}$ induced by the $\tril$-dependent diagram of Fig.~\ref{fig1} can be written as 
\beq \left(\delta \mathcal{M}
\right)_{Z_{H}} = \left( \sqrt{Z_{H}} -1 \right) \mathcal{M}^0,~~~~~~~~ Z_H = \frac1{1- \ktre^2\, \delta Z_H}\,,
\label{delzh}
\eeq   
where $\mathcal{M}^0$  is the lowest-order amplitude  and 
\beq
\delta Z_{H} = -\frac{9}{16} \, 
\frac{G_\mu\,\mh^2 }{\sqrt{2}\, \pi^2} \left(\frac{2\pi}{3\sqrt{3}}-1\right)\,.
\label{deltaZH}
\eeq
 
In order to extend the range of convergence of the perturbative expansion to large values of  $\ktre$,
the one-loop contribution in $Z_H$ has been resummed. In so doing, terms of $\ord((\ktre^2 \alpha)^n)$ which are expected to be the dominant higher-order corrections at large $\ktre$ are correctly accounted for. 

In addition to the  ${\tril}^2$ universal term above, amplitudes depend linearly on $\tril$ differently for each process and kinematics. Let $\mathcal{M}^0$ be the Born amplitude corresponding to a given process (production or decay). 
At the level of cross section or decay width, the linear dependence on $\tril$ originates from the interference of the Born amplitude $\mathcal{M}^0$ and the virtual EW amplitude $\mathcal{M}^1$,  besides the wave function renormalisation constant.  The amplitude $\mathcal{M}^1$ involves one-loop
diagrams when the process at LO is described by tree-level diagrams,
like, {\it e.g.}, vector boson fusion production, while it involves
two-loop diagrams when the LO contribution is given by one-loop
diagrams, like, {\it e.g.}, gluon-gluon-fusion production.  The
$\tril$-linearly-dependent contributions in $\mathcal{M}^1$, which we
denote as $\mathcal{M}^1_{\tril}$, can be obtained for any process by
evaluating in the SM  the diagrams that contain one trilinear
Higgs coupling ($\mathcal{M}^1_{\trilsm}$) and then rescaling them by
a factor $\ktre$. 
In order to correctly identify $\mathcal{M}^1_{\trilsm}$ (the
contributions related to the $H^3$ interaction) in the
$\mathcal{M}^1$ amplitude in the SM, it is convenient to choose a
specific gauge, namely the unitary gauge. In a renormalisable
$R_\xi$ gauge, $\trilsm$-dependent diagrams are due not only to the
interaction among three physical Higgs fields but also to the
interaction among one physical Higgs and two unphysical scalars, making
the identification less straightforward.

Once all the contributions from $\mathcal{M}^1_{\tril}$ and $Z_H$ are
taken into account, denoting as $\Sigma$ a generic cross section for
single Higgs production or a Higgs decay width, the corrections
induced by an anomalous trilinear coupling  modify the LO
prediction ($\Sigma_{\rm{LO}}$) according to
\beq
\Sigma_{\rm{NLO}} = Z_H\,\Sigma_{\rm{LO}} \left( 1 + \ktre C_1 \right)\,,
\label{NLOEWSM}
\eeq   
where the coefficient $C_1$, which originates from  $\mathcal{M}^1_{\trilsm}$, 
depends on the process and the kinematical observable considered, while $Z_H$ is universal, see Eq.~\eqref{delzh}.  Here and in the following the LO contribution is understood as including QCD corrections so that the labels LO
and NLO refer to EW corrections. We remind that among all terms contributing to the complete EW corrections  we  consider only the part relevant for our discussion, {\it i.e.}, the one related to the Higgs trilinear interaction.  
The $\Sigma_{\rm{NLO}}$ in the SM  can be  obtained from
Eq.~(\ref{NLOEWSM}) setting $\ktre =1$ and expanding the $Z_H$ factor,
or \beq \Sigma^{\rm SM}_{\rm{NLO}} = \Sigma_{\rm{LO}} \left( 1 + C_1 +
\delta Z_H \right)\,.
\label{SigmaSM}
\eeq
Thus,  the relative corrections induced by an anomalous trilinear Higgs 
self-coupling can be expressed as 
\beq
\dsigma \equiv 
\frac{\Sigma_{\rm{NLO}} - \Sigma^{\rm SM}_{\rm{NLO}}}{\Sigma_{\rm{LO}}} =
Z_H -(1 + \delta Z_H) + (Z_H \ktre -1 ) C_1 \, ,
\label{corr}
\eeq
which, neglecting $\ord(\ktre^3\,\alpha^2)$  terms in the r.h.s, can be compactly written as
\beq
\dsigma = (\ktre-1) C_1+(\ktre^2-1) C_2 \,,
\label{correxp}
\eeq
with
\beq
C_2 = \frac{\delta Z_H}{(1- \ktre^2 \delta Z_H)}~.
\label{eqC2}
\eeq  
Before describing the method and results of the calculation of the $C_1$ coefficients, we scrutinise
the theoretical robustness of Eq.~$\eqref{corr}$ and its range of
validity.  Our aim is to employ Eq.~$\eqref{corr}$ to evaluate the
LHC sensitivity on $\tril$ without making ``a priori''  any assumptions on the value of the parameter $\ktre$.
We will, however, demand as a consistency constraint that, for large values of $\ktre$,
$\tril$-dependent terms from $\ord(\alpha^{j})$ corrections with $j>1$ do 
not overwhelm the effects from the $C_i$ coefficients. In order to take into account all
the $\ord((\ktre^2 \alpha)^n)$ contributions and perform a resummation of the  $\ktre^2\,\delta Z_{H}$ terms in $Z_H$ we  need to impose that $\ktre^2\,\delta Z_{H} \lesssim 1$, {\it i.e.}, $|\ktre| \lesssim 25$.
The corresponding parametric uncertainty in $\Sigma_{\rm{NLO}}$ is therefore given by 
$\ord((\ktre^3 \alpha^2))$ terms that can be sizeable for large
values of $\ktre$. The size of such missing terms can be estimated by calculating 
the difference between $\dsigma$ computed using  Eq.~(\ref{corr}) and Eq.~(\ref{correxp}), or equivalently
$\delta (\Sigma_{\rm{NLO}}/\Sigma_{\rm{LO}}) \simeq \ktre^3 C_1 \delta Z_H$. 
Requiring  this uncertainty to be $ \lesssim 10\%$ and assuming as an order of magnitude of 
the two-loop contribution $ C_1 \delta Z_H \sim 10^{-5}$,
we find $|\ktre| \lesssim 20$, which we take as the range of validity of our perturbative calculation.

\medskip

At variance with the SM, where the Higgs self coupling and the Higgs mass are related, 
in our setup they are two independent parameters. This in general spoils the renormalisability of the 
model and makes its parameters sensitive to the UV scales. However, one knows a priori that the $\tril$--dependent $\ord(\alpha)$ corrections to $\Sigma$ in Eq.~\eqref{corr} are finite. The reason is twofold: \\

i) the LO  result does not  depend on $\tril$ and therefore no renormalisation of $\tril$ at NLO is either needed nor possible.\\

ii)  All the counterterms needed at NLO do not contain divergent contributions proportional to the trilinear coupling. \\

This last point can be understood as follows: the only counterterm that contains divergent contributions proportional to $\tril$  is the Higgs mass counterterm. However, the $\mh$ dependence  in $\Sigma_{\rm LO}$ is all of kinematical
origin. Therefore, when the NLO corrections are calculated, no renormalisation of $\mh$ is needed. 

The arguments above are sufficient for all the processes except for $H \to \gamma \gamma$, which deserves a dedicated discussion. In a $R_\xi$ gauge the LO dependence of $\Gamma (H \to \gamma \gamma)$ upon
$\mh$ is not purely kinematical, but it also comes from diagrams containing
unphysical charged scalars. Therefore one expects that in these gauges
at NLO there is no clear way to disentangle the contributions that can be 
assigned as due to a trilinear coupling from the ones related to the kinematical
parameter $\mh$. In order to overcome this difficulty, as we already said, we 
employed the unitary gauge. In this gauge all the LO $\mh$ dependence of  
$\Gamma (H \to \gamma \gamma)$ is  kinematical, similarly to  all the other 
observables we considered, and the argument discussed above about the
finiteness of the  NLO $\tril$--dependent  corrections applies.

\medskip 

In general, an anomalous coupling $c_i$ is a free parameter that does not 
satisfy the SM relations  that can be crucial for the renormalisability of the model. 
In the calculation of radiative corrections, the substitution of an electroweak coupling with
an anomalous one, $c_i^{\rm SM}\TO c_i \equiv \kappa_{i} c_i^{\rm SM}$  gives 
a  finite result in two cases. First, when the renormalisation of $c_i$ 
does not involve EW corrections. Second, when the renormalisation of the other regular
couplings $c_j$ involves $c_i$ via EW corrections, but $c_i$ itself is
not renormalised.  The first case corresponds to what happens in the context of the $\kappa-$formalism
where couplings are rescaled by overall factors. It also applies to
many phenomenological and experimental studies on the dependence of double Higgs
 production cross sections on $\tril$ as done, {\it  e.g}, in \cite{Baglio:2012np} or in the experimental studies
\cite{ATL-PHYS-PUB-2014-019, ATL-PHYS-PUB-2015-046}. In this case only QCD
higher-order corrections can be consistently included. The second case
corresponds to the study presented here: $\Sigma$ at LO does not
depend on $\tril$ and  the NLO EW corrections, which do depend on $\tril$,
are finite because do not involve the renormalisation of $\tril$. At this point, it is worth stressing that
studies analogous in spirit and philosophy to ours have been performed for the case of the top-Higgs Yukawa coupling $y_t$, where, by looking at the dependence of NLO EW corrections, bounds on 
anomalous $y_t \equiv \kappa_t y_t^{\rm SM}$ can be set via the
analysis of top-quark pair production measurements \cite{Kuhn:2013zoa,Beneke:2015lwa}.

It should be said that, while the $\ord(\asa{i}{})$ corrections to the
physical observables $\Sigma$ due to an anomalous trilinear Higgs
coupling are finite, and therefore they do not provide us with direct
information about the scale $\Lambda$ of NP, one expects that the
$\ord(\asa{i}{j})$ corrections with $j>1$ will instead show at least a
logarithmic sensitivity to $\Lambda$. For our analysis to be
trustworthy, one has to therefore make the further assumption that the
scale $\Lambda$ is not too far from the EW scale, such that
potentially large logarithmic corrections that would spoil the
perturbativity of our analysis are not there.

In summary, we have argued that loop-induced dependence of single
Higgs processes on $\tril$ can be seen in the same spirit as, for
example, the dependence of Higgs pair production cross sections on
$\tril$ or the general fits of the anomalous Higgs couplings at the
LHC in the $\kappa$-framework. The variable $\ktre$ in
Eq.~\eqref{corr} is a parameter that can be directly probed at the
experimental level, looking for discrepancies from SM predictions. The
value of $\ktre$ is {\it a priori} unconstrained, besides the limits
imposed by perturbativity; constraints on its value can be set via
experimental data. Clearly, if an UV-completed BSM model is specified
or an EFT approach is used then different range of validity should be
set on the parameter $\ktre$.

Finally, let us stress that our investigation probes a larger range of
$\ktre$  with respect to an Effective-Field-Theory approach based on
the addition of the dimension six operator $(\Phi^\dagger \Phi)^3$, as proposed 
for instance in Ref.~\cite{Gorbahn:2016uoy}. In this case the requirement that 
the potential is bounded from
below and $v$ being the absolute minimum  sets the constraints $1<\ktre<3$ as 
shown in Appendix \ref{sec:EFT}.

\section{Computation  of the $C_1$ coefficients }
\label{sec:coefficients}

At variance with the $C_2$ coefficient, which is universal, the $C_1$
coefficients are process- and kinematic-dependent and therefore need
separate calculations. In this work we focus on the main production
and decay channels:
\begin{itemize}
\item $\sigma_{\ggF}$, the gluon-gluon-fusion cross section;
\item $\sigma_{\text{VBF}}$, the VBF cross section;
\item $\sigma_{WH}$, $\sigma_{ZH}$, the cross section for associate production with $W$ and $Z$ bosons;
\item $\sigma_{\tth}$,  the cross section for $\tth$ production;
\item $\Gamma_{\gamma\gamma}$,  the decay width into photons;
\item $\Gamma_{ZZ}$ and $\Gamma_{WW}$,  the decay widths into $ZZ^*$ and $WW^*$ subsequently 
decaying into fermions;
\item $\Gamma_{f\bar{f}}$, the decay width into fermions;
\item $\Gamma_{gg}$, the decay width into gluons.
\end{itemize}

For each observable, the corresponding $C_{1}$ coefficient is identified as the
contribution linearly proportional to $\trilsm$ in the NLO EW
corrections and normalised to the LO result as evaluated in the SM. 

For any given single Higgs process, in principle $C_{1}$ could be evaluated directly at the level of matrix element in a fully differential way, {\it i.e.}, point by point in phase space
\beq
C_1(\{p_n\})=
\frac{ 2\Re(\mathcal{M}^{0*} \mathcal{M}^1_{\trilsm})}
{ |\mathcal{M}^{0}|^2}\, ,
\label{C1me}
\eeq where we have explicitly shown in parentheses the dependence on
the external momenta $\{p_n\}$ in the Born configuration and
understood the sum/average over helicities and colour states. By
integrating over the phase space the differential ratio in
Eq.~(\ref{C1me}) one would achieve the maximal discriminating power
between the $\ktre=1$ hypothesis and the $\ktre \neq 1$ ones. However,
as first step, it is both useful and convenient to work at the more
inclusive level and directly compute $C_1$ for cross sections or decay
rates integrated over the entire phase space or a portion of it.

For example, in the case of the decays, in this work we limit the discussion to total rates and define $C_1^\Gamma$ as 
\beq
C_1^\Gamma=
\frac{\int d\Phi~ 2\Re \left( \mathcal{M}^{0*}\mathcal{M}^1_{\trilsm} \right)}
{\int d\Phi~  |\mathcal{M}^{0}|^2}\, ,
\label{C1f}
\eeq
where the integration in $d\Phi$ is over the phase space of the final-state particles.

The computation of (total or differential) hadronic cross sections is more involved than that of the decay widths, because
hadronic cross sections receive contributions from different partonic process, which have to be convoluted with the corresponding parton luminosities and in principle can have different $C_1$ terms at the level of matrix elements.
For production cross section, $C_1^\sigma$ reads 
\beq
C_1^\sigma=
\frac { \sum_{i,j}  \int dx_1 dx_2 f_i(x_1) f_j(x_2) \,  2\Re \left (\mathcal{M}_{i j}^{0*}\mathcal{M}^1_{\trilsm,i j} \right) d\Phi 
 }  { \sum_{i,j}   \int dx_1 dx_2 f_i(x_1) f_j(x_2)  \,  |\mathcal{M}_{i j}^{0}|^2  d\Phi  }  \, ,
\label{C1h}
\eeq
where the sum is over  all the possible $i,j$ partonic initial states  of the 
process, which are convoluted with the corresponding parton distribution functions.

A few comments on the $C_1$ for the various observables considered
here are in order before showing the results. 
Assuming that all the fermions but the top quark are massless,  the $C_1^\Gamma$ for $H\TO Z Z^* \TO 4f$ does not depend on the fermions in the final state. The same applies to $H\TO W W^* \TO 4f$. 
In the case of hadronic production,  different partonic processes can have different
$C_1$'s at the level of matrix elements. One example is $t \bar t H$ production, which receives contributions
from $q\bar q\TO t \bar t H$ and $gg \TO t \bar t H$. Another is VBF, where both $W$-boson-fusion and 
$Z$-boson-fusion contribute. Moreover, each subprocess contributes in proportion to the parton distribution weights.

\begin{figure}[t]
\begin{center}\vspace*{-1.0cm}
\includegraphics[width=0.49\textwidth]{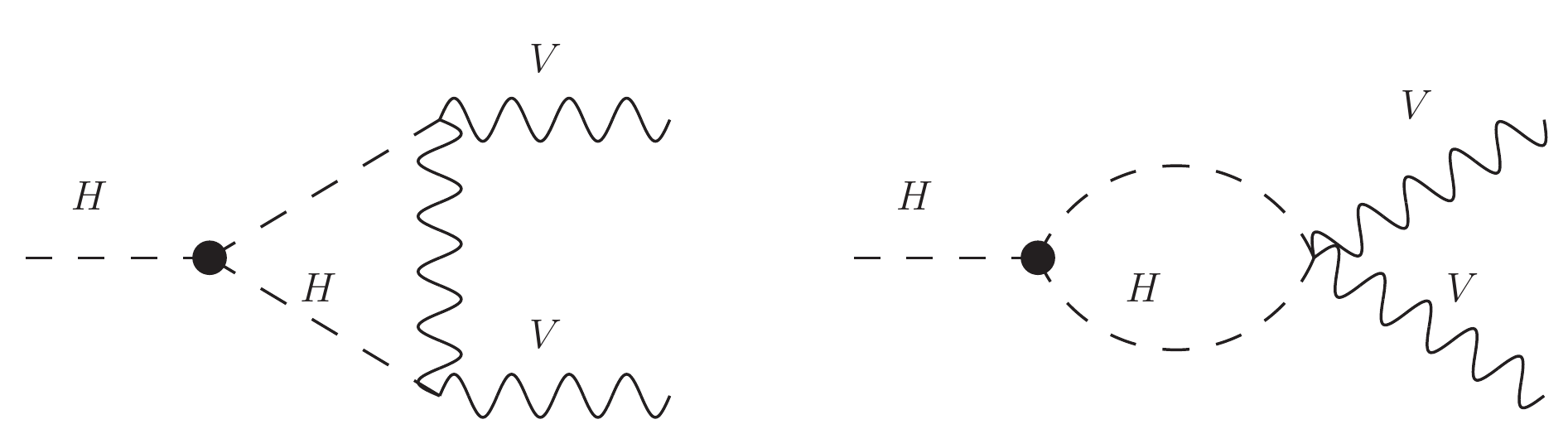}
\caption{Structure of the $\trilsm$-dependent part in  $\mathcal{M}^1_{\trilsm}$
for processes involving massive vector bosons in 
the final or in the intermediate states (VBF, $HV$ and $H\TO V V^* \TO 4f$).}
\label{fig2}
\end{center}
\end{figure}

\begin{figure}[t]
\begin{center}
\includegraphics[width=0.3\textwidth]{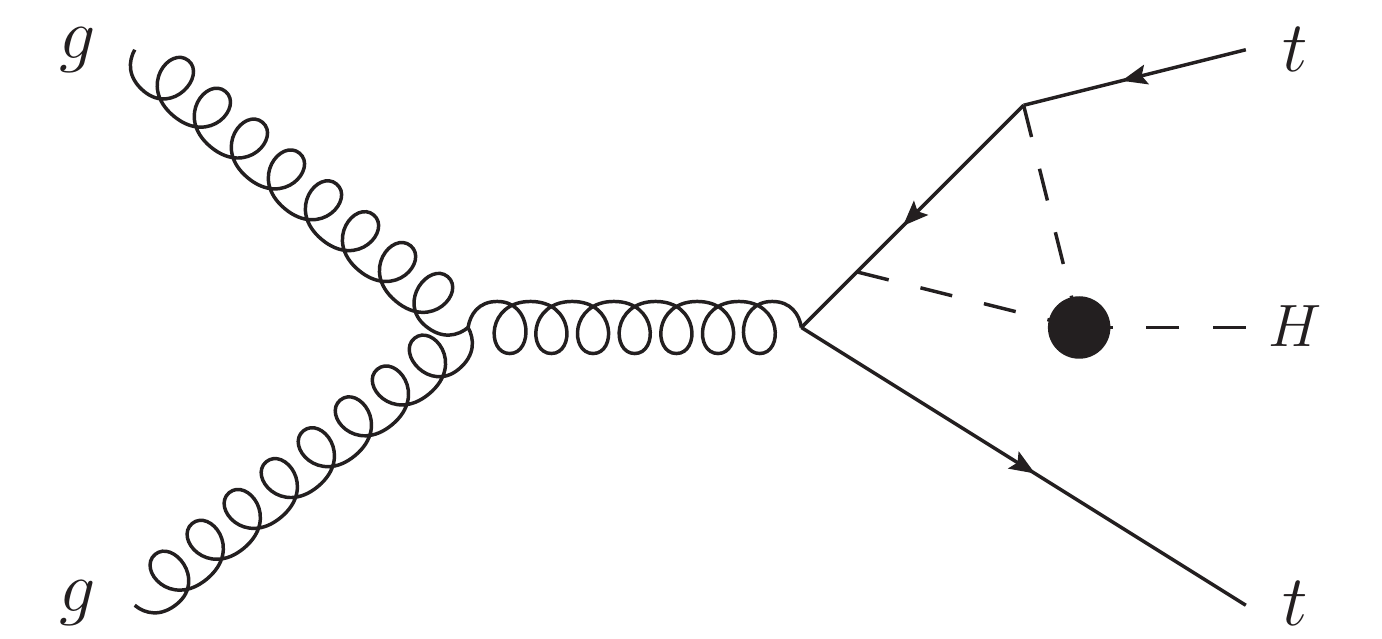}
\includegraphics[width=0.3\textwidth]{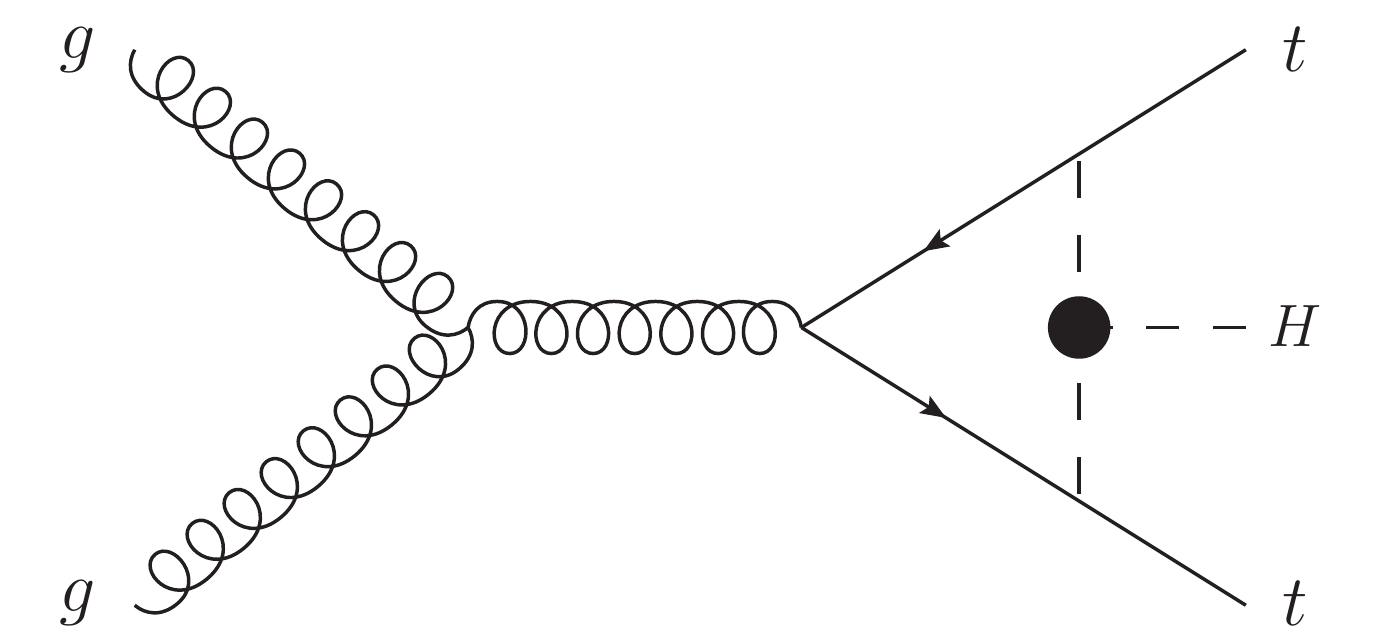}
\includegraphics[width=0.3\textwidth]{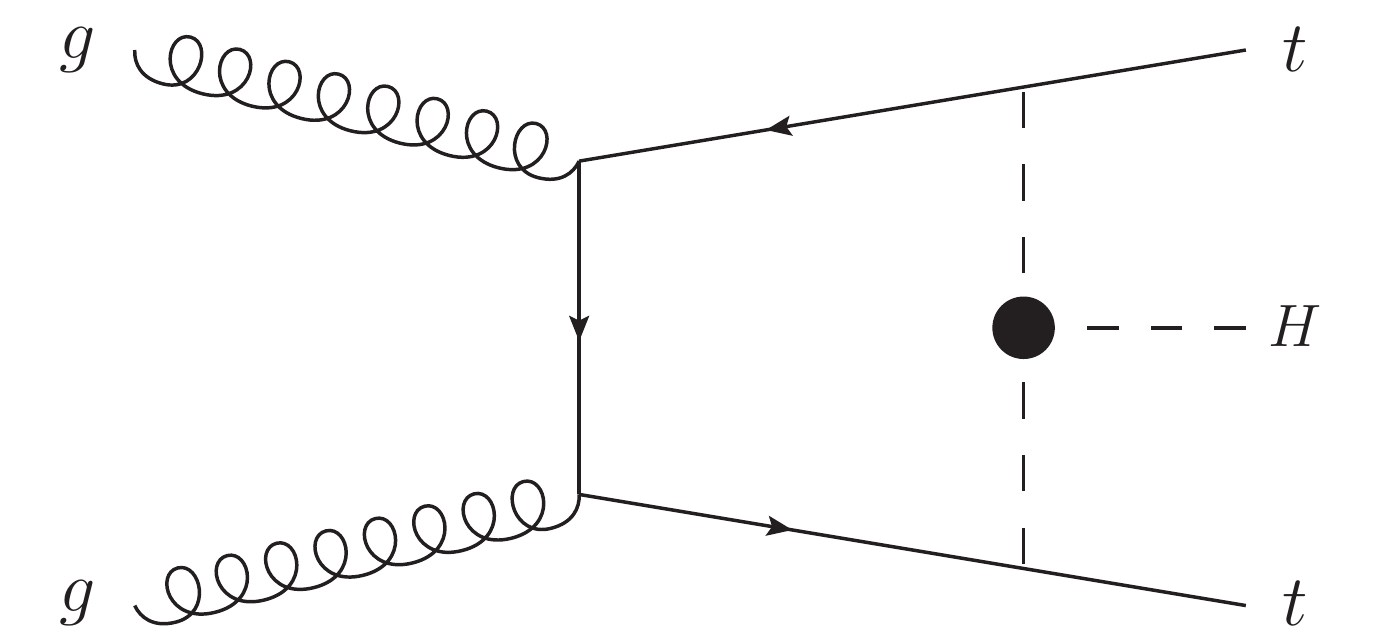}
\caption{ Sample of $\trilsm$-dependent diagrams in $t\bar{t}H $
  production.  }
\label{fig20}
\end{center}
\end{figure}

In order to evaluate the $C_1$ coefficients of the various processes,
we generated the relevant amplitudes using the Mathematica package
{\sc\small FeynArts} \cite{Hahn:2000kx}. For all the cases involving
only one-loop amplitudes, we computed the cross sections and decay
rates with the help of {\sc\small FormCalc} interfaced to {\sc\small
  LoopTools} \cite{Hahn:1998yk} and we checked the partonic cross
sections at specific points in the phase space with FeynCalc
\cite{Mertig:1990an,Shtabovenko:2016sxi }. In processes involving
massive vector bosons in the final or in the intermediate states (VBF,
$HV$ and $H\TO V V^* \TO 4f$), the $\tril$-dependent parts in
$\mathcal{M}^1_{\trilsm}$ have a common structure, see
Fig.~\ref{fig2}. In the case of the $t\bar{t}H$ production the
sensitivity to $\tril$ comes from the one-loop corrections to the
$t\bar{t}H$ vertex and from one-loop box and pentagon diagrams.  A
sample of diagrams containing these $\tril$-dependent contributions is
shown in Fig.~\ref{fig20}.

The presence of not only triangles but also boxes and pentagons in the
case of $t \bar t H$ production provides an intuitive explanation of
why the $\lambda_3$ contributions cannot be captured by a local
rescaling of the type that a standard $\kappa$-framework would assume
for the top-Higgs coupling. Similarly, not all the contributions given
by the corrections to the $HVV$ vertex can be described by a scalar
modification of its SM value via a $\kappa_V$ factor, due to the
different Lorentz structure at one loop and at the tree level.

\begin{figure}[t]
\begin{center}
\includegraphics[width=0.30\textwidth]{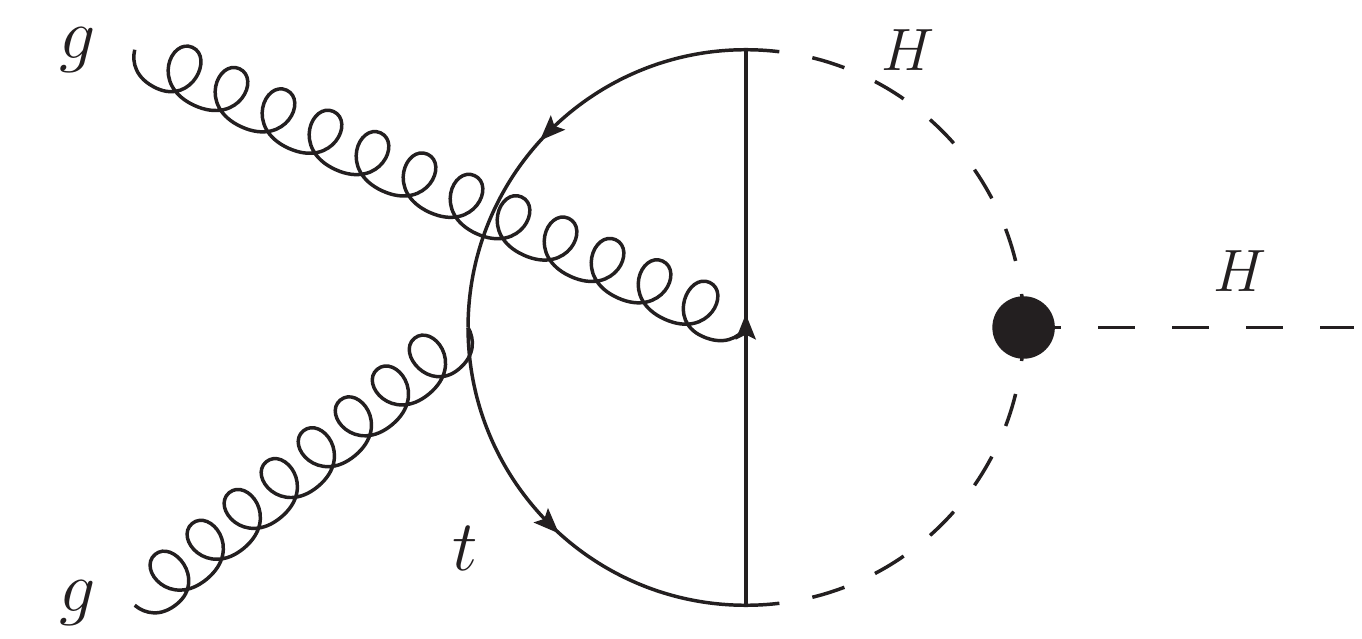}
\includegraphics[width=0.37\textwidth]{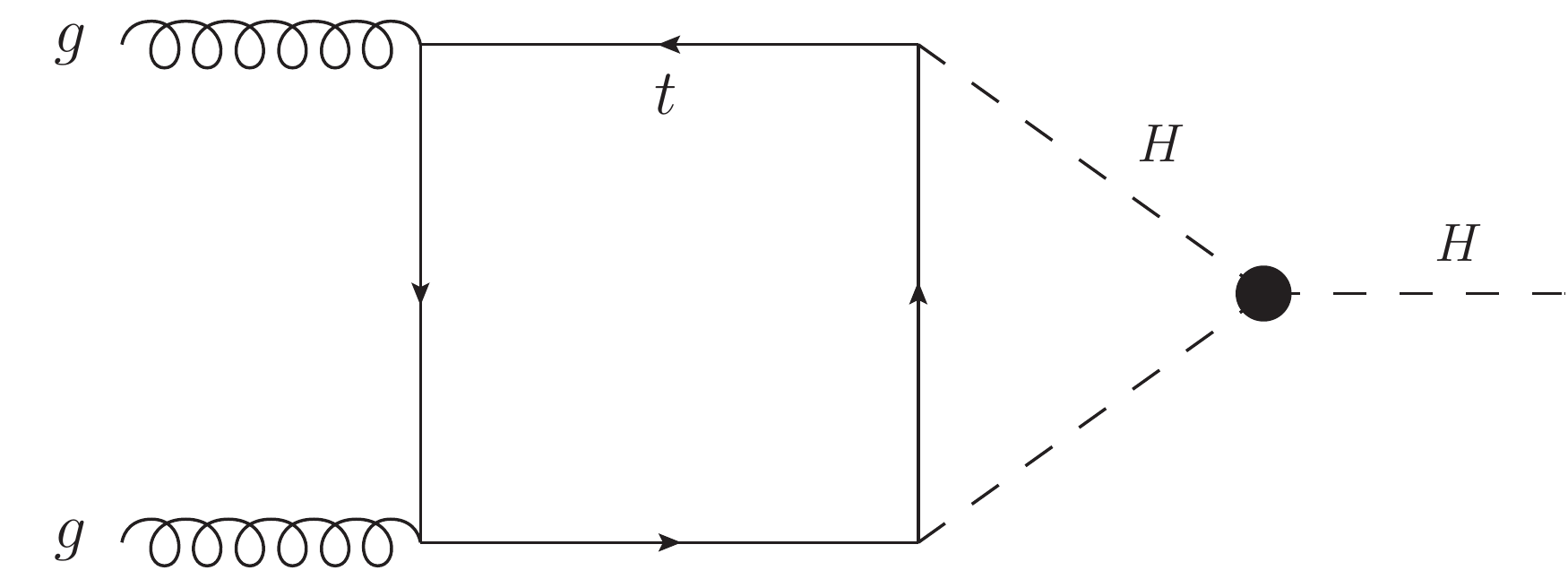}
\caption{Diagrams contributing to the $C_1$ coefficient in the gluon-gluon-fusion
Higgs production. The one on the right has a multiplicity factor 2.}
\label{fig3}
\end{center}
\end{figure}

The computation of $\sigma(gg\rightarrow H)$, the related $\Gamma(H
\to gg)$, and of $\Gamma(H \rightarrow\gamma\gamma)$ is much more
challenging and deserves a more detailed discussion. These observables
receive the first non-zero contributions from one-loop diagrams, which
do not feature $\lambda_3$, so that the computation of $C_1$ requires
the evaluation of two-loop diagrams.

The two-loop EW corrections to $\sigma(gg\rightarrow H)$ in the SM
were obtained in
Refs.\cite{Aglietti:2004nj,Degrassi:2004mx,Actis:2008ug}.  In our
computation of the $C_1$ coefficient we followed the approach of
Ref.~\cite{Degrassi:2004mx} where the corrections have been computed
via a Taylor expansion in the parameters $q^2/(4 \mt^2), \: q^2/(4
\mh^2)$ where $q^2$ is the virtuality of the external Higgs momentum,
to be set to $\mh^2$ at the end of the computation.  However, at
variance with Ref.\cite{Degrassi:2004mx}, we computed the diagrams
contributing to $C_1$, see Fig.~\ref{fig3}, via an asymptotic
expansion in the large top mass up to and including ${\cal
  O}(\mh^6/\mt^6)$ terms. The two expansions are equivalent up to the
first threshold encountered in the diagrams that defines the range of
validity of the Taylor expansion. In our case, the first threshold in
the diagrams of Fig.~\ref{fig3} occurs at $q^2 = 4\, \mh^2$ and both
expansions are valid for $\mh \simeq 125$ GeV. The asymptotic
expansion was performed following the strategy described in
Ref.\cite{Degrassi:2010eu} and the result for $C_1$ is presented in
Appendix \ref{app1}. We checked our asymptotic expansion against the
corresponding expression obtained by the Taylor expansion finding, as
expected, very good numerical agreement.

\begin{figure}[t]
\begin{center}
\includegraphics[width=0.30\textwidth]{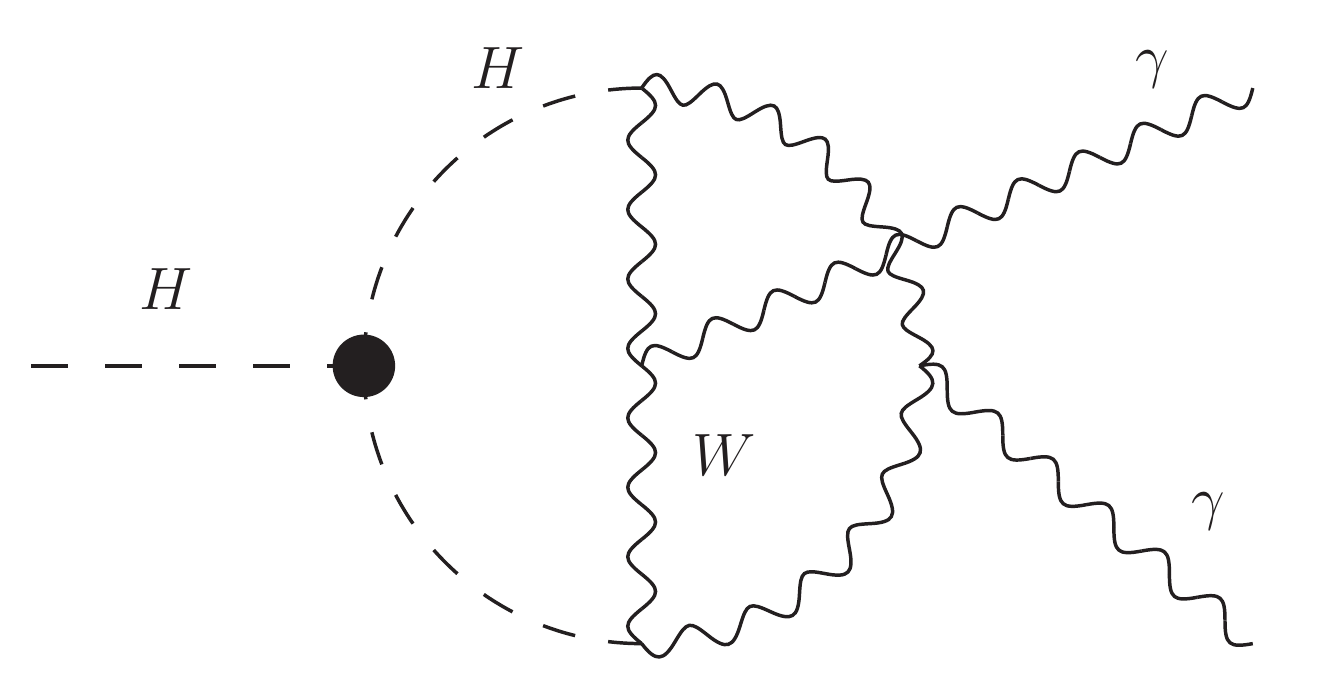}
\includegraphics[width=0.30\textwidth]{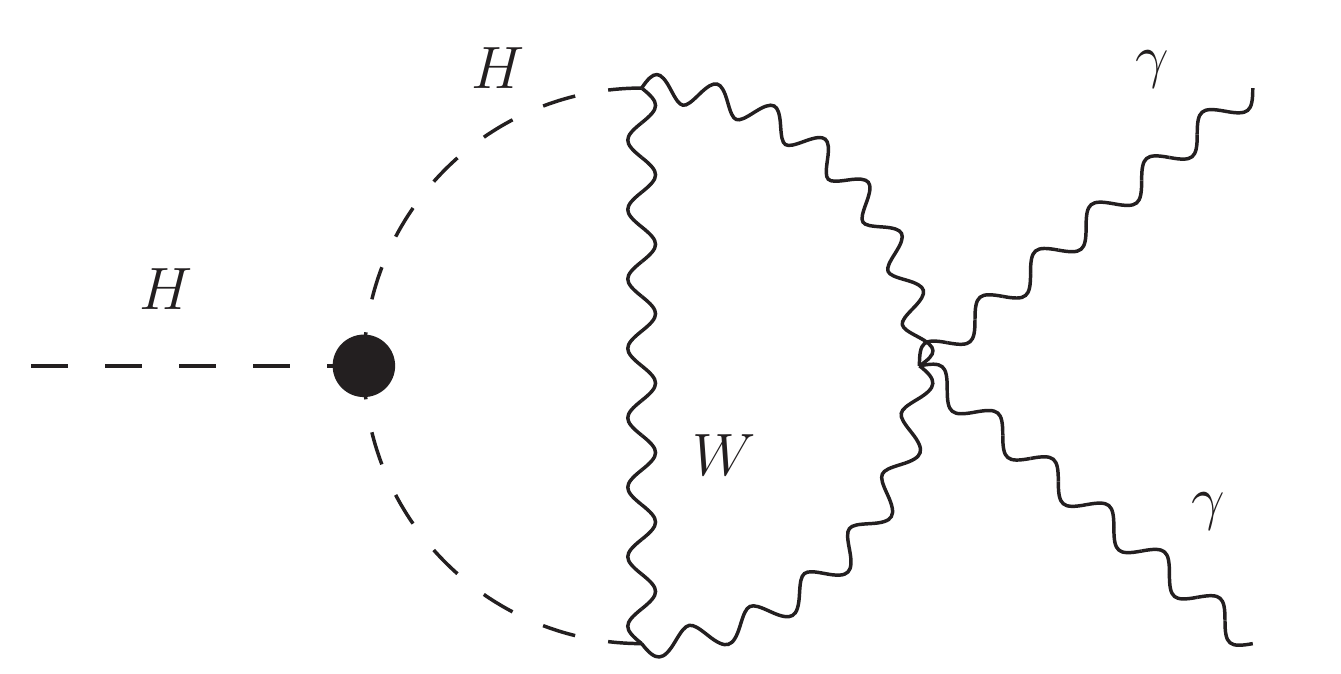}
\includegraphics[width=0.38\textwidth]{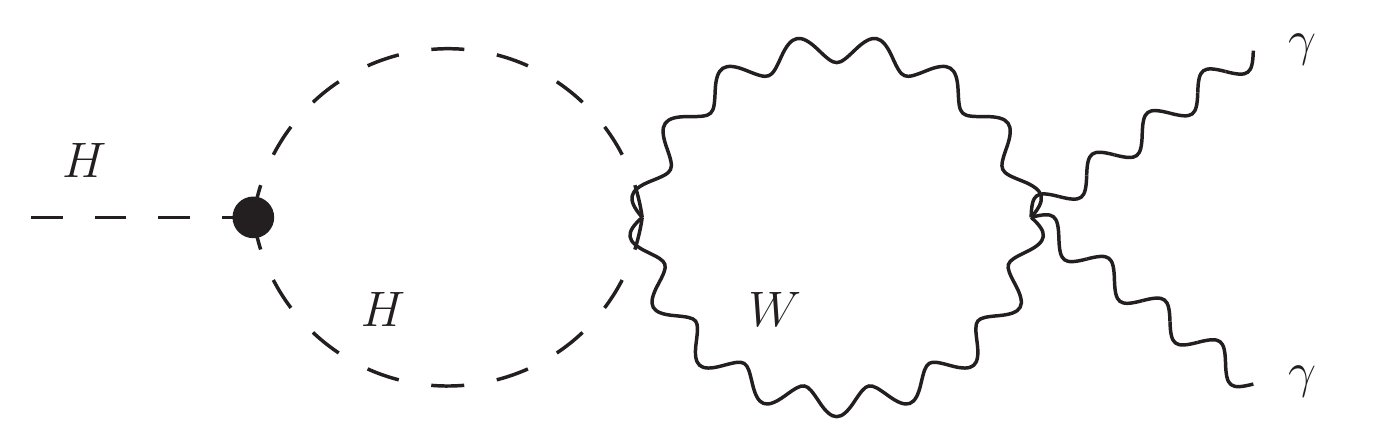}
\includegraphics[width=0.35\textwidth]{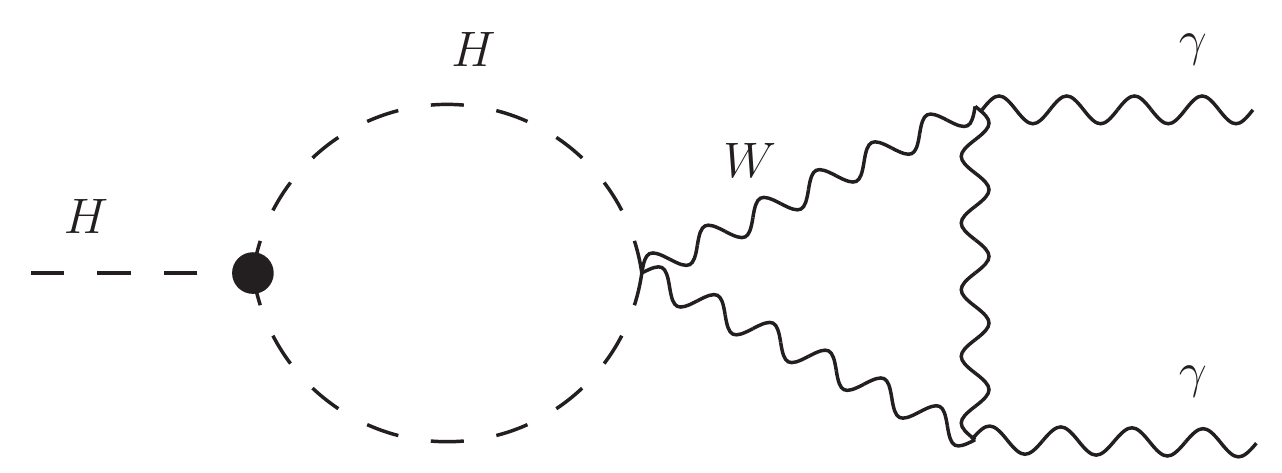}
\includegraphics[width=0.35\textwidth]{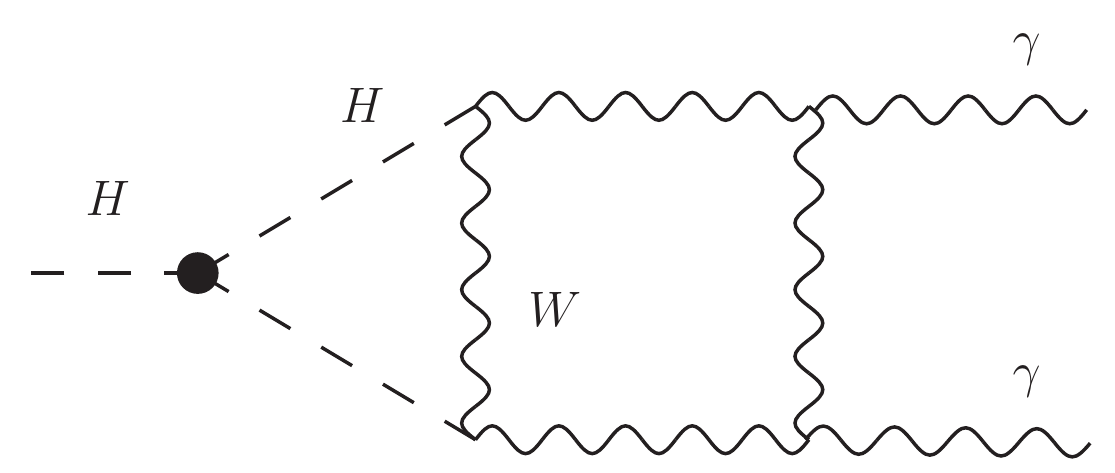}
\caption{Diagrams contributing to the $C_1$ coefficient in 
$\Gamma ( H \to \gamma \gamma)$. The diagrams in the second row have 
multiplicity 2. }
\label{fig4}
\end{center}
\end{figure}

The computation of the EW corrections to the partial decay width of a
Higgs boson into two photons in the SM was performed in a $R_\xi$
gauge in Refs.\cite{Degrassi:2005mc,Actis:2008ts}.  As mentioned
above, the identification of the contributions to the $C_1$
coefficient is straightforward in the unitary gauge. In this gauge,
neither unphysical scalars nor ghosts are present and the propagator
of the massive vector bosons is $i (- g_{\mu \nu} + k_\mu
k_\nu/M_V^2)/(k^2 - M_V^2 + i\epsilon)$.  The unitary gauge is a very
special gauge.  It can be defined as the limit when the gauge
parameter $\xi$ is sent to infinity of a $R_\xi$ gauge. When a
calculation is performed in the unitary gauge, one is actually
interchanging the order of the operations limit $\xi \to \infty$ with
the integration, {\it i.e.}, the limit $\xi \to \infty$ is performed
first and then one does the integration while the correct order is the
opposite. Because some of the vertices that arise from the
gauge-fixing function contain a $\xi$ factor, this exchange is not
always an allowed operation and in order to check the correctness of
our approach we recomputed\footnote{To our knowledge this is the
  first-ever two-loop computation of a physical observable performed
  in the unitary gauge.}  the full two-loop EW corrections to $\Gamma
(H \to \gamma \gamma)$ in the unitary gauge.  The corrections were
computed as in Ref.\cite{Degrassi:2005mc} via a Taylor expansion in
the parameters 
$q^2/(4 \mt^2), \: q^2/(4 \mw^2), \: q^2/(4 \mh^2)$ up to and including
${\cal O}(q^6/m^6)$ terms finding perfect agreement with the result of
Ref.\cite{Degrassi:2005mc}.

Once we verified that in the SM the calculation in the unitary gauge is
equivalent to the one in a $R_\xi$ gauge, the coefficient $C_1$ is obtained 
evaluating  the  diagrams in the unitary gauge  that contain one trilinear 
Higgs interaction. The latter  amounts to  add to the contribution of the
diagrams in Fig.~\ref{fig3}, with the gluons replaced by photons,  the
contribution of the diagrams in Fig.~\ref{fig4}. The result is presented
in Appendix \ref{app1}. We would like to remark that 
the sum of the diagrams in Fig.~\ref{fig4}   is finite  in the
unitary gauge but it is not finite in a generic $R_\xi$ gauge.

 \section{Results}
\label{sec:results}
In this section we discuss the numerical impact of the
$\tril$-dependent contributions on the most important observables in single Higgs production and decay at the LHC. 
We begin by listing and commenting the size of the $C_{1}$ and $C_{2}$ factors in Eq.~(\ref{correxp}), which
parametrise the $\tril$-dependent contributions.

The input parameters of our calculation are \cite{MelladoGarcia:2150771} 
\begin{equation}
G_{\mu}=1.1663787 \cdot 10 ^{-5} ~\gev^{-2}\,,\; \mw = 80.385 ~\gev\,,\;  \mz = 91.1876 ~\gev\,, \label{EWinput}
\end{equation}
with the Higgs boson and the top-quark masses set to  
\begin{equation}
m_{H} = 125 ~\gev\,,\;  \mt = 172.5 ~\gev \label{mhmt}\, .
\end{equation}
All the other fermions are treated as massless. In the production
cross sections, the renormalisation and factorisation scales are both
set equal to
\begin{equation}
\mu \equiv \frac{1}{2} \sum_{i} m_{i}  \label{scale}\,,
\end{equation}
where $m_{i}$ are the masses of the particle in the final state. As
PDF set, we use the PDF4LHC2015 set \cite{Butterworth:2015oua,Dulat:2015mca,Harland-Lang:2014zoa,Ball:2014uwa}.

The process-independent factor $C_2$ defined in Eq.~(\ref{eqC2})
depends upon $\delta Z_H$, as defined in Eq.~(\ref{deltaZH}), and
also $\ktre$. With the parameter inputs used, $\delta Z_H =-1.536
\cdot 10^{-3}$, thus $C_2$ can range from $C_2 = -1.536 \cdot 10^{-3}$
for $\ktre = 1$ up to $C_2 = -9.514 \cdot 10^{-4}$ for $\ktre = \pm
20$.

\begin{table}[t]
\begin{center}
\begin{tabular}{l|l|l|l|l|l}
$C_1^{\Gamma}[\%]$	& ${\gamma\gamma}$ &${ZZ}$& ${WW}$&  ${f\bar{f}}$ &${gg}$\\
\hline
on-shell $H$  &0.49 & 0.83& 0.73 & 0 & 0.66
\end{tabular}
\end{center}
\caption{Values of the $C_1$ factor in units $10^{-2}$ for the most relevant decay modes of the Higgs boson.}
\label{c1g}
\end{table}

\begin{table}[t]
\begin{center}
\begin{tabular}{l|l|l|l|l|l}
$C_1^{\sigma}[\%]$	& ${\ggF}$ &
${\text{VBF}} $ &  ${WH}$ & ${ZH}$ &
${\tth}$\\
\hline
7 TeV &0.66 & 0.65  &1.06 &1.23 & 3.87 \\
8 TeV &0.66 & 0.65 &1.05 & 1.22 & 3.78 \\
13 TeV &0.66 & 0.64  &1.03 & 1.19 & 3.51 \\
14 TeV &0.66 & 0.64  &1.03 & 1.18 & 3.47 \\
\end{tabular}
\end{center}
\caption{Same as Tab.~\ref{c1g}  for the production modes for $pp$ collisions at centre-of-mass energies relevant for the LHC.}
\label{c1s}
\end{table}

In Tab.~\ref{c1g} we report the values of the $C_1^{\Gamma}$ term for
the most relevant Higgs decay modes at the LHC, namely, $WW$, $ZZ$, $\gamma \gamma$, $f \bar f$ and also $gg$, which yields a non-negligible fraction
of the total decay width. In the analyses of section \ref{sec:limits},
$C_1^{\Gamma}(f\bar f)=0$ is used for the $b \bar b$ and $\tau
\tau$ decays.  The $C_1^{\sigma}$ factors for the different
single Higgs production modes are presented in Tab.~\ref{c1s} for different 
centre-of-mass energies of  Run-I and Run-II at the LHC.
For all the processes, the scale uncertainty obtained by scaling $\mu$
with a factor of 2 and 1/2 amounts to 1\% of the value displayed in
Tab. \ref{c1s}. The dependence on the factorisation scale
largely cancels in the ratio of Eq.~\ref{C1h} and the
dependence on the renormalisation scale is either not present ($VH$,
VBF) or also cancels exactly in the ratio. 

Few comments can be given about the results in Tabs.~\ref{c1g} and
\ref{c1s}.  The term $C_1^\Gamma(f \bar f)$ is proportional to $m_{f}$
for a generic $H\TO f\bar f$ fermionic decay. We have verified that in the
case of $H\TO b\bar b$, setting $m_b=5~\gev$, $C_1^\Gamma(b \bar
b)=2.5\times10^{-5}$. Thus it is safe to set $C^{\Gamma}_1(f \bar f)$ for any $H\TO f\bar f$ (and in particular for $C^{\Gamma}_1(b \bar b)$) 
decay to zero. The smallest non-vanishing $C_1$  corresponds to the 
$H\TO \gamma\gamma$ channel. 
It is interesting to note that, besides
subleading kinematical effects, the main difference in the
determination of $C_1^\Gamma(ZZ)$ and $C_1^\Gamma(WW)$ is the
different coupling of the Higgs boson with the gauge bosons in
Fig.~\ref{fig2}. For this reason, $C_1^\Gamma(ZZ)/C_1^\Gamma(WW)\sim
m_Z / m_W $ and similarly $C_1^\sigma(ZH)/C_1^\sigma(WH)\sim m_Z / m_W
$. On the other hand, $C_1^\Gamma(ZZ)$ is different form
$C_1^\sigma(ZH)$, although the vertex corrections involved are the
same (see Fig.~\ref{fig2}). In this case the kinematic
configurations are not the same, leading to different 
values for $C_1^\Gamma(ZZ)$ and $C_1^\sigma(ZH)$. A similar argument applies
to $C_1^\Gamma(WW)$ and $C_1^\sigma(WH)$ and for a comparison
with $C_1^\sigma({\rm VBF})$. 

Another interesting observation that can be drawn from Table~\ref{c1s}
regards the dependence of $C_1^{\sigma}$ from the hadronic
centre-of-mass energy, which, although it is very mild for all
processes, points to the fact that the effects become smaller at
higher energies. Furthermore, we note that the $\tth$ production
receives much larger corrections with respect to the other processes,
while Higgs-strahlung processes, $ZH$ and $WH$, receive larger
corrections than VBF and gluon-gluon-fusion. The behaviour with energy
and the hierarchy can be nicely understood by considering the
Yukawa-type potential induced by the Higgs interaction in the
non-relativistic regime.\footnote{Similar effects have been discussed,
  {\it e.g.}, in the case of $t \bar t$ production in
  \cite{Kuhn:2013zoa}.}  In $\tth$, $WH$ and $ZH$ production the Higgs
can interact with another final-state particle via an Higgs
propagator, thus in the non-relativistic regime the process receives a
Sommerfeld enhancement. On the contrary, this is not possible in
gluon-gluon-fusion, VBF and in the decays into $\gamma\gamma$ and
$ZZ(WW)$, where the $\mathcal{M}^1_{\trilsm}$ involves always a Higgs
propagator connecting the external Higgs with an internal line.  This
explains why, although the interactions are the same, $C_1^{\sigma}(t
\bar t H)> C_1^{\sigma}(\ggF)$ and also $C_1^{\sigma}(HV)>
C_1^{\sigma}(\rm{VBF}),C_1^{\Gamma}(VV)$.

In order to support the arguments outlined above, the kinematical
dependence of the $C_1$ coefficients can be studied.  To this purpose,
we evaluate $C_1^{\sigma}$ for these processes imposing an upper cut
on the transverse momentum of the Higgs or on the total invariant mass
of the final state. The results obtained for 13-TeV collisions are
shown in Tabs.~\ref{c1kinpt} and \ref{c1kinminv}, for the cases
$p_T(H)<p_{T, {\rm cut}}$ and $m_{\rm tot}<K\cdot m_{\rm thr}$, being
$m_{\rm thr}$ the threshold of the specific process.  $C_1^\sigma$ is
strongly enhanced when energetic configurations are vetoed. In this
respect, boosted configurations, which feature a smaller cross section
and a milder dependence on $\ktre$, are certainly not optimal to
detect deviations in the Higgs trilinear coupling. On the other hand,
the selection of threshold regions may improve the sensitivity on
$\ktre$.  Results for VBF have not been included in the table because
the dependence on the cuts turns out to be very mild (very few
percentages with respect to the value in table \ref{c1s}), as expected
from the fact that the $\lambda_3$ dependence involves $HVV$ vertex
corrections, which are not connected with the quark lines.
 
\begin{table}[t]
\begin{center}
\begin{tabular}{l|l|l|l|l|l}
$C_1^{\sigma}[\%] $
& $25~\gev$ &
$50~\gev$ &  $100~\gev$ & $200~\gev$ &
$500~\gev$\\
\hline
$WH$ &1.71 (0.11) & 1.56 (0.34) &1.29 (0.72) &1.09 (0.94) & 1.03 (0.99)\\
$ZH$ &2.00 (0.10) & 1.83 (0.33)&1.50 (0.71)& 1.26 (0.94) & 1.19 (0.99)\\
$t \bar t H$ &5.44 (0.04) & 5.14 (0.17)  &4.66 (0.48)& 3.95 (0.84)& 3.54 (0.99)\\
\end{tabular}
\end{center}
\caption{$C_1^\sigma$ at 13 TeV obtained by imposing the cut $p_T(H)<p_{T, {\rm cut}}$, for several values of  
$p_{T, {\rm cut}}$. In parentheses the fraction of events left after the quoted cut is applied.}
\label{c1kinpt}
\end{table}

\begin{table}[t]
\begin{center}
\begin{tabular}{l|l|l|l|l|l}
$C_1^{\sigma}[\%] $
& $ ~~~~~1.1$ &
$~~~~~1.2$ &  $~~~~~1.5$ & $~~~~~2$ &
$~~~~~3$\\
\hline
$WH$ &1.78 (0.17) & 1.60 (0.36) &1.32 (0.70) &1.15 (0.89) & 1.06 (0.97)\\
$ZH$ &2.08 (0.19) & 1.86 (0.38)&1.51 (0.72)& 1.31 (0.90) & 1.22 (0.98)\\
$t \bar t H$ &8.57 (0.02) & 7.02 (0.10)  &5.11 (0.43)& 4.12 (0.76)& 3.64 (0.94)\\
\end{tabular}
\end{center}
\caption{$C_1^\sigma$ at 13 TeV obtained by imposing the cut $m_{\rm tot}<K\cdot m_{\rm thr}$, for several values of $K$.  In parentheses the fraction of events left after the quoted cut is applied.}

\label{c1kinminv}
\end{table} 
 
\medskip

We turn now to the presentation and discussion of the results for
production and decay.  We first consider the corrections $\dsigmah$ to
the various channels as defined in Eq.~\eqref{corr}.
\begin{figure}[t]
\centering
\includegraphics[width=0.475\textwidth]{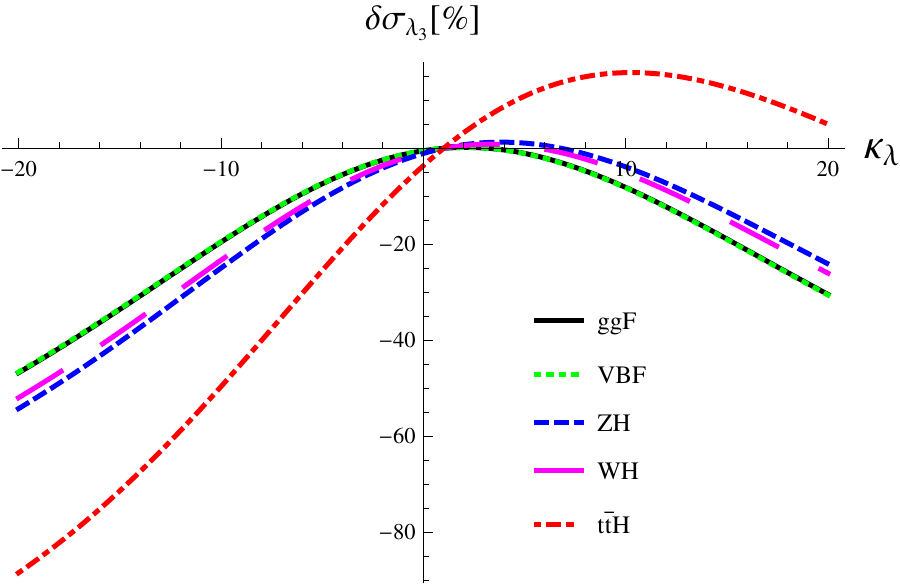}
\includegraphics[width=0.475\textwidth]{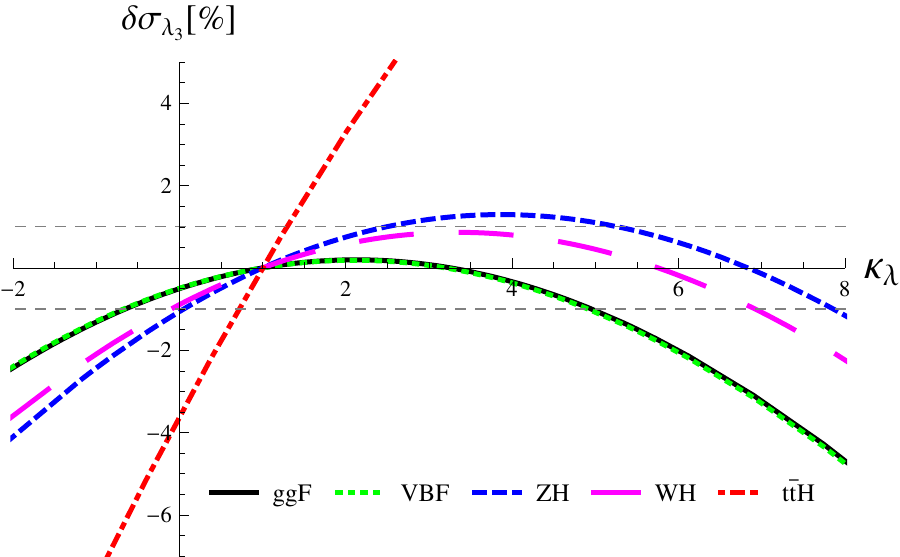}
\caption{Dependence of $\dsigmah$ for the relevant production processes at the 
LHC as a function of $\ktre$ in the range $|\ktre|\leq 20$ (left)  and zoomed in the region $-2 < \ktre <8$  (right).
The style and colour conventions of the lines are: 
$\ggF$ = solid black,  $t \bar t H$ = dash-dotted red,  VBF = dotted green, 
$ZH$ = dashed blue, $WH$ = long-dashed magenta. The black dashed horizontal lines in the right plot correspond to $\pm1\%$.}
\label{fig:corrprods_with_n2}
\end{figure}
In Fig.~\ref{fig:corrprods_with_n2} we plot $\dsigmah$ as a function
of $\ktre$ for the relevant production processes at the LHC, namely,
gluon--gluon fusion, vector-boson-fusion, Higgs-strahlung ($WH$ and
$ZH$) and $t \bar t H $ production.  In the plot on left we display
the $\dsigmah$ corrections for the various processes in the full range
of validity of our calculation, $-20 \lesssim \ktre \lesssim 20$,
while in the plot on the right we zoom the region $-2<\ktre<8$, where
corrections are within $5\%$ in absolute value for all processes but
$\tth$.

As can be seen, $\tth$ receives positive sizeable corrections ($\sim20
\%$ at $\ktre\sim 10$), thanks to the large value of
$C_1^{\sigma}(\tth)$. For all the other production processes large
corrections can only be negative and only for large value of
$|\ktre|$. The plots on the right of Fig.~\ref{fig:corrprods_with_n2}
shows that $\dsigmah$ remains at the percent level for a quite
extended range for the $\ggF$, VBF and $VH$ production
modes. Moreover, for these processes, $\dsigmah$ can be zero for
values of $\ktre\neq 1$, {\it i.e.}, different from the SM
prediction. In particular, in the case of gluon-gluon fusion and VBF,
the SM is degenerate with $\ktre\sim3$, while in the case of $VH$
production the SM is degenerate with $\ktre\sim6$. The fact that the
degeneracy appears at different values $\ktre$ for different processes
is important in order to be able to lift it.

\begin{figure}[t]
\centering
\includegraphics[width=0.45\textwidth]{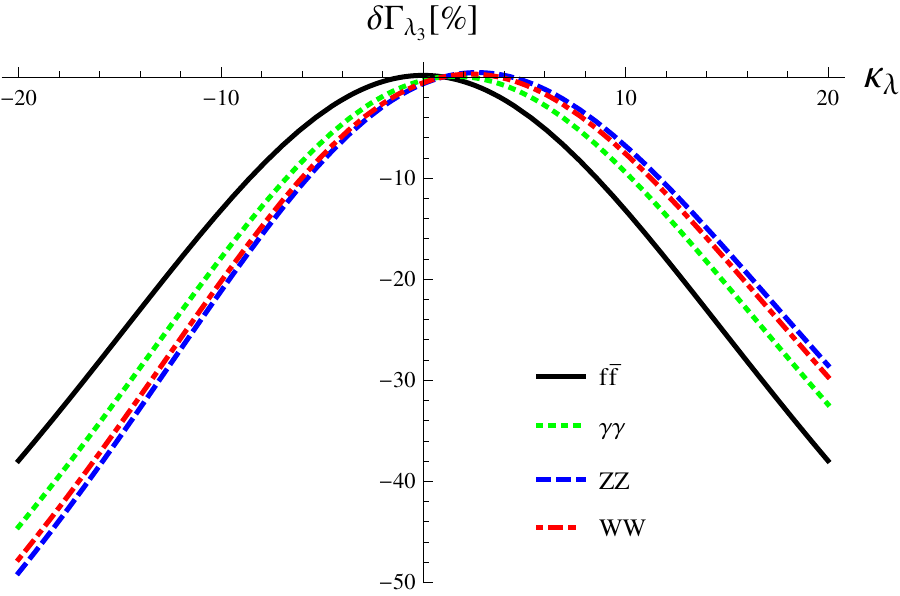}
\includegraphics[width=0.45\textwidth]{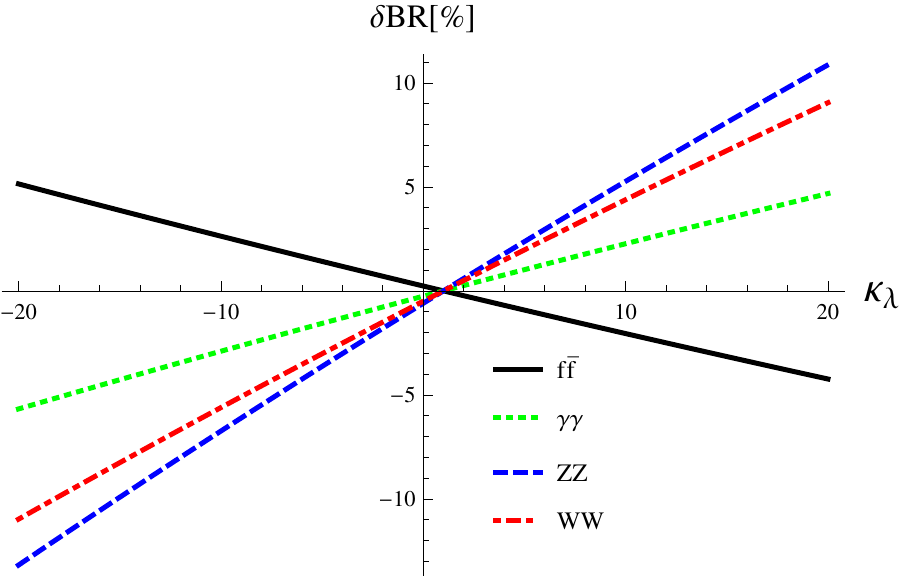}
\caption{
Dependence of $\dgamma$ for the relevant decay widths (right)
and corresponding $\dBR$ as defined in Eq.~(\ref{BRform}) (left). The solid
black line represents $\Gamma_{f \bar f}$, the long-dashed red line $\Gamma_{\sss WW}$,
the dashed blue line  $\Gamma_{\sss ZZ}$ and the dotted green line  $\Gamma_{\gamma \gamma}$.}
\label{fig:corrdecays_with_n2}
\end{figure}

The results for the decay widths and branching ratios are shown Fig.~\ref{fig:corrdecays_with_n2}. 
We plot (left) $\dsigma$ as a function of $\ktre$ for the decay widths of the relevant
modes at the LHC, which we denote as $\dgamma$, and we show (right) the analogous quantity ($\dBR$) for the Branching Ratios (BRs).  The quantity $\dBR(i)$ for the Higgs decay into the final-state $i$ can be conveniently written as
\beq
\dBR (i) =  
                      \frac{  (\ktre-1) (C^{\Gamma}_1(i)- C_1^{\Gamma_{\rm tot}})}
                     {1+ (\ktre-1) C_1^{\Gamma_{\rm tot}}} \,,
\label{BRform} 
\eeq
where we have defined $C_1^{\Gamma_{\rm tot}}\equiv \,\sum_{j}   \br^{\rm SM} (j)  C^{\Gamma}_1(j)$ and with our input parameters $C_1^{\Gamma_{\rm tot}}= 2.3\cdot 10^{-3}$.  The quantity $C_1^{\Gamma_{\rm tot}}$, which actually is the $C_1$ term for the total decay width, is very small since $C^{\Gamma}_1(b \bar b)=0$ and $b \bar b$ is the dominant decay channel. Note that, although the $H\TO gg$ decay is not phenomenologically relevant, the total decay width does depend on $\dgamma(gg)$, since $\Gamma_{gg}$ yields a non-negligible fraction (8.5 \%) of $\Gamma_{\rm tot}$.

Figure~\ref{fig:corrdecays_with_n2} shows that the corrections to the
partial widths can reach up to $-40\%$ or $-50 \%$ for $\ktre \sim -20$, while  for $\ktre > 0$ the 
corrections are smaller due to the different sign of the contributions depending on $C^{\Gamma}_1$ and $C_2$.
The only exception is $ \delta\Gamma_{\tril}({f \bar f})$, which is symmetric since $C^{\Gamma}_1(f \bar f)$=0. 
On the other hand, the corrections to the branching ratios $\dBR$, which are more important than  $ \delta\Gamma_{\tril}$ from a phenomenological point of view, are much smaller, reaching up to $\sim10 \%$ for $\br(ZZ)$. The reasons behind the smallness of the $\dBR$ are two. First, as  explicitly shown in Eq.~(\ref{BRform})   $\dBR$  depends only linearly upon $\ktre$, since 
the contribution of the wave function renormalisation constant cancels in the ratio. Second, the $C_1$ coefficients have the same sign and therefore there is a partial cancellation in the ratio. In any case, it is interesting to note that in the range of $\ktre$ shown in the right-hand plot of Fig.~\ref{fig:corrprods_with_n2}, apart from $t \bar t H$, the terms $\dBR$ are of the same size or larger than $\dsigmah$.  In other words, in the range close to the SM predictions, the decays modes  are  more sensitive to $\ktre$ than the production processes.
\\

\section{Constrains on $\tril$: present and future}
\label{sec:limits}

In this section we describe the method and the results of a simplified fit we have performed in order to 
estimate the limits that can be set on $\ktre$ with our approach. Our analysis is based on the experimental results presented in Tab.~8 of Ref.~\cite{Khachatryan:2016vau}.  We also estimate the expected limits that could be obtained at LHC Run-II at 300 fb$^{-1}$ and 3000 fb$^{-1}$ of luminosity. 

The key aspect of our approach is that the predictions for all the available production and decay channels 
depend on a single parameter ($\kappa_\lambda$) and
therefore a global fit can be in principle very powerful in
constraining the Higgs trilinear coupling. As our aim is mostly illustrative, 
we want to assess the competitiveness of our method rather than trying
to obtain the best and most robust bounds. To this purpose, we make a series of simplifying approximations. 
For example, being usually quite small (see Fig. 7 of Ref.~\cite{Khachatryan:2016vau}),
we ignore  correlations between the different uncertainties of a single
measurement or between the measurements of the different observables. 

The basic inputs of our analysis are the signal strength parameters $\muif$, which
are defined for any specific combination of production and decay
channel $i \to H \to f$ as \beq \muif \equiv \mu_i \times \mu^f =
\frac{\sigma(i)}{\sigma(i)^{\rm SM}} \times \frac{\br(f)}{\br^{\rm
    SM}(f)}~.
\label{signalstre}
\eeq 
The quantities $\mu_i$ and $\mu^f$ are  the
production cross section $\sigma(i)$ 
($i=$ $\ggF$, {\rm VBF}, $WH$, $ZH$, $t \bar tH$) and the $\br(f)$ $(f= \gamma\gamma, ZZ, WW, b\bar{b},
\tau\tau)$ normalised to their SM values, respectively. Assuming on-shell
production, the product $\mu_i \times \mu^f$ is therefore the rate for the $i \to H \to f$ process 
normalised to the corresponding SM prediction.

Using Eq.~(\ref{corr}) and Eq.~(\ref{BRform}), $\mu_i$ and $\mu^f$, which enter the definition of  $\muif$ in Eq.~(\ref{signalstre}), 
can be expressed as
\beqn 
\mu_i &=& 1 + \dsigmah(i) \,, \nonumber \\
\mu^f &=& 1 + \dBR(f) \,.
\label{eq:muf}
\eeqn
By definition, $\muif=\mu_i=\mu^f=1$ in the SM. 

In the following we denote the measured signal strengths as $\muifexp$.
Given a collection of $\muifexp$ measurements $\{ \muifexp \}$, we define as  best value of $\ktre$  the one that minimises the $\chi^2(\ktre)$ function defined as
\begin{equation}\label{chi2}
\chi^2(\ktre)\equiv 
 \sum_{\muifexp \in \{ \muifexp \}} \frac{(\muif(\ktre)-\muifexp)^2}{(\Delta_i^f (\ktre))^2}\,,  
\end{equation}
where 
 $\muif(\ktre)$ is obtained using Eqs.~(\ref{signalstre}) and (\ref{eq:muf}), and $\Delta_i^f(\ktre)$ is the total uncertainty of $\muif$. Different sources of uncertainties enter in the determination of $\Delta_i^f(\ktre)$, namely, the experimental uncertainty in the measurement of $\muif$, the SM theory uncertainties associated to the particular channel  $\mu_i \times \mu^f$ (scale, PDFs and $\alpha_s$),  and the $\ktre$-dependent uncertainty associated to missing higher orders, the $\ord(\ktre^3 \alpha^2)$ terms discussed in Sec.~\ref{sec:org}. 
The first two types of uncertainty are reported already combined in Ref.~\cite{Khachatryan:2016vau}, and divided in experimental and theoretical errors in Ref.~\cite{Peskin:2013xra}.
For the third type of uncertainty, we adopt the parametrization $\frac1{\sqrt{3}}\ktre^3 C_1 \delta Z_{H}$, where the $C_1$ depends on the observable and $\delta Z_{H}$ is defined in Eq.~(\ref{deltaZH}). It has to be  kept in mind, however, that the results of our analysis show a very mild dependence on this uncertainty.~\footnote{The prefactor $1/\sqrt{3}$ is included so that the uncertainty very closely corresponds to the difference between Eq.~\ref{corr} and   Eq.~\ref{correxp}. }
\begin{table}[t]
\begin{center}
\begin{tabular}{llllllllll|l|}
& $H\rightarrow\gamma\gamma$& $H\rightarrow ZZ$ & $H\rightarrow W W $&$ H\rightarrow\tau\tau$ & $H\rightarrow b \bar b$\\
\hline
$\ggF$ & P$_{1,2,3,4}$; F$_{1,2}$ & P$_{1,2,3,4}$; F$_{1,2}$& P$_{1,2,3,4}$; F$_{1,2}$& P$_{1,2,3,4}$; F$_{1}$ & ---\\
\hline
VBF &P$_{2,3,4}$;  F$_{1,2}$ & P$_{2,3,4}$;  F$_{1,2}$ & P$_{2,3,4}$;  F$_{1,2}$ & P$_{2,3,4}$;  F$_{1,2}$  & --- \\
\hline
$WH$ & P$_{3,4}$ &---&P$_{3,4}$ & P$_{3,4}$&  P$_{3,4}$;F$_{1,2}$ \\
\hline
$ZH$ & P$_{3,4}$& ---& P$_{3,4}$& P$_{3,4}$&P$_{3,4}$ \\
\hline
$t \bar tH$ & P$_{4}$; F$_{1,2}$ &---& P$_{4}$& P$_{4}$&  P$_{3,4}$;F$_{1,2}$ \\
\end{tabular}
\end{center}
\caption{Combinations of production and decay modes used in the different 
analyses. Each P$_{n}$ identifies one of our four different sets of present 
data taken from Ref.~\cite{Khachatryan:2016vau}. F$_{1}$ and F$_{2}$ 
respectively correspond to the future scenarios  ``CMS-II'' (300 fb$^{-1}$) 
and  ``CMS-HL-II'' (3000 fb$^{-1}$) as presented in Tab.~1 of 
Ref.~\cite{Peskin:2013xra}.}
\label{tab:exp}
\end{table}

\begin{figure}[t]
\centering
\includegraphics[width=0.45\textwidth]{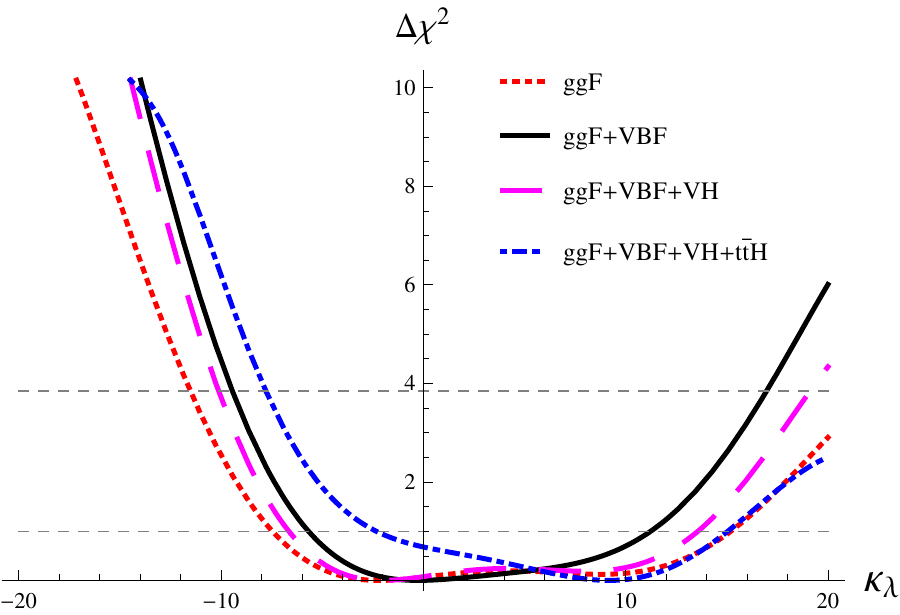}
\includegraphics[width=0.45\textwidth]{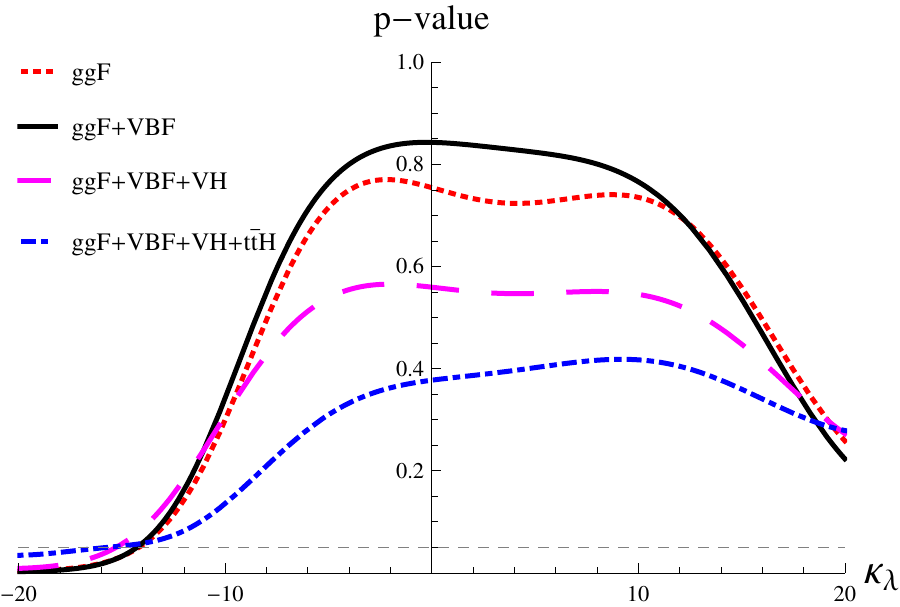}
\caption{Left: $\chi^2$ for the different sets of observables presented in Tab.~\ref{tab:exp}: the dotted red line represents P$_{1}$, the solid black line P$_{2}$, the dashed magenta line P$_{3}$, and the blue dash-dotted line P$_{4}$. The two horizontal lines represent $\Delta \chi^2=1$ and $\Delta \chi^2=3.84$. Right: corresponding $p$-value. The various P$_{n}$ data sets are colour-coded in the same way. The horizontal line is $p=0.05$.}
\label{fig:chisq}
\end{figure}

In order to evaluate the impact of the different  production channels on the fit to the present data, we consider four different sets (P$_{n}$), with an increasing number of included production channels: 
\begin{itemize}
\item P$_{1}$: $\ggF$,
\item P$_{2}$: $\ggF$+VBF, 
\item P$_{3}$: $\ggF$+VBF+$VH$, 
\item P$_{4}$: $\ggF$+VBF+$VH$+${t\bar t H}$.
\end{itemize}
 For the future scenarios (F$_{n}$), we consider 
  \begin{itemize}
 \item F$_{1}$:  ``CMS-II'' (300 fb$^{-1}$),
 \item F$_{2}$: ``CMS-HL-II'' (3000 fb$^{-1}$), 
 \end{itemize}
 as presented in Tab.~1 of Ref.~\cite{Peskin:2013xra}.
  A summary of the sets of data used in each fit is presented in Tab.~\ref{tab:exp}.

As shown in Fig.~\ref{fig:chisq}, we identify the $1\sigma$ and
$2\sigma$ intervals assuming a $\chi^2$ distribution.  Following this
procedure and using the gluon-gluon-fusion and VBF data from Tab. 8 of
Ref.~\cite{Khachatryan:2016vau} (scenario P$_{2}$ in
Tab. \ref{tab:exp}) we obtain \beqn \ktre^{\rm best}=-0.24\,, ~~~~~~
\ktre^{1 \sigma} = [-5.6,11.2]\,, ~~~~~~ \ktre^{2 \sigma} =
     [-9.4,17.0]\,,
\label{muf}
\eeqn 
where the $\ktre^{\rm best}$ is the best value and $\ktre^{1
  \sigma}$, $\ktre^{2 \sigma}$ are respectively the $1\sigma$ and
$2\sigma$ intervals. The choice of P$_{2}$ as reference set is
motivated by the measured significance for the different production
processes, which in the 8 TeV analyses is  above $5\sigma$ only for $\ggF$ and
VBF (see Tab.~14 in Ref.~\cite{Khachatryan:2016vau}). Moreover,
P$_{2}$ returns the most stringent values for $\ktre^{1 \sigma}$ and
$\ktre^{2 \sigma}$. The other data sets presented in
Tab.~\ref{tab:exp} are reported in Fig.~\ref{fig:chisq}.  Notice how
the minimum of the distribution in the figure jumps to $\sim$ 10 when
the $t\bar{t} H$ production channel is included. This effect
originates from the anomalous values presented in
Ref.~\cite{Khachatryan:2016vau} for $\bar{\mu}_{\tth}^f$, especially
with $f=WW$. Similarly, the low compatibility of $\bar{\mu}_{VH}^f$
with SM predictions is the reason behind larger $\ktre^{1 \sigma}$ and
$\ktre^{2 \sigma}$ intervals in P$_{3}$.

In order to ascertain the goodness of our fit, we computed the $p$-value as a function of $\ktre$:
\beqn 
\text{$p$-value}(\ktre)=1-F_{\chi^2_{(n)}}(\chi^2(\ktre))\,,
\eeqn
where $F_{\chi^2_{(n)}}(\chi^2(\ktre))$ is the cumulative distribution function for a $\chi^2$ distribution with $n$ degrees of freedom, computed at $\chi^2(\ktre)$. In the right-hand side of Fig.~\ref{fig:chisq} we report the $p$-value$(\ktre)$ corresponding to different data sets. Requiring that $p>0.05$, we are able to exclude, at more than $2\sigma$, that a model with an anomalous coupling $\ktre<-14.3$ can explain the data  in P$_{2}$. 

\begin{figure}[t]
\centering
\includegraphics[width=0.45\textwidth]{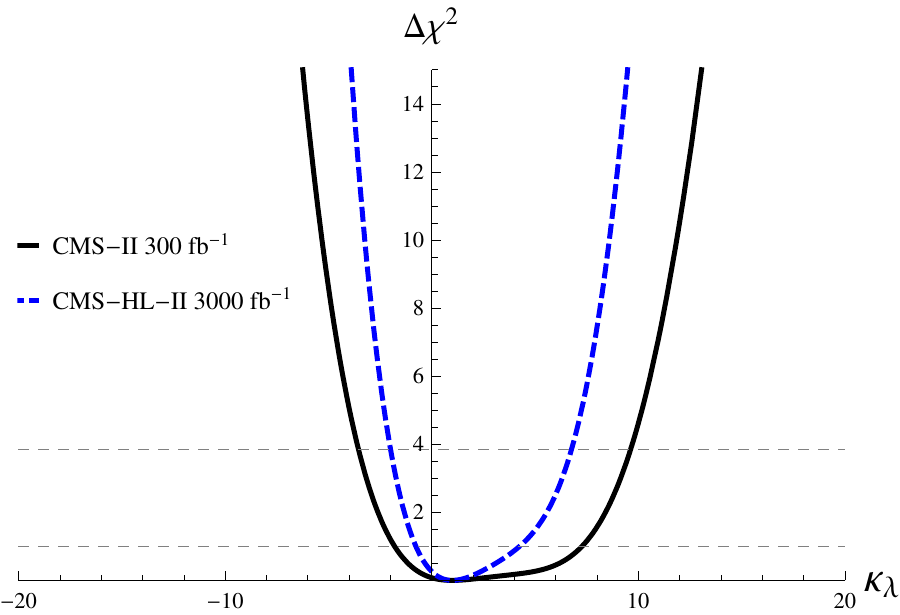}
\includegraphics[width=0.45\textwidth]{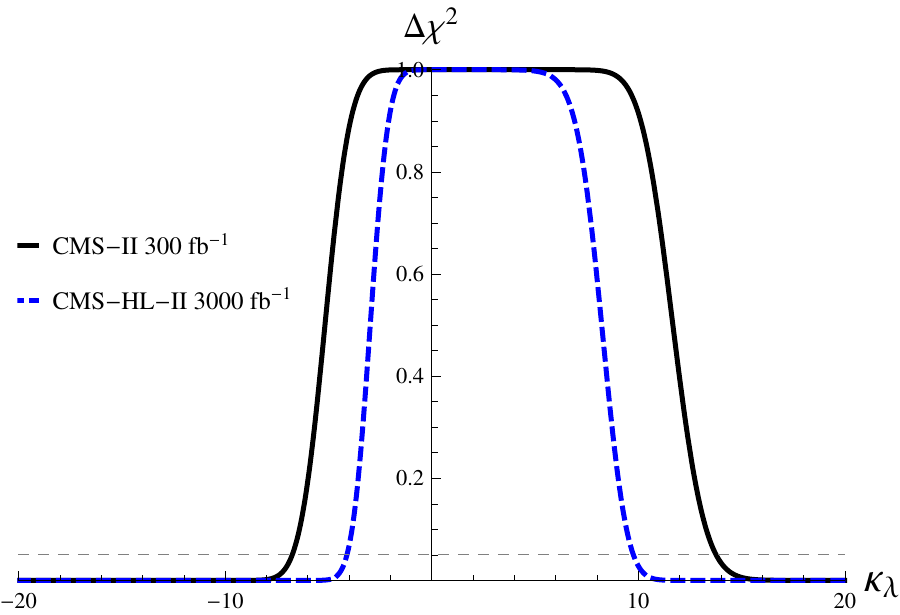}
\caption{In the left and right plots, respectively $\chi^2(\ktre)$ and $p$-value$(\ktre)$ for ``CMS-II'' (solid black line) and ``CMS-HL-II'' (blue dashed line)}
\label{Chifuture}
\end{figure}

We repeat the same procedure for ATLAS and CMS at 300 fb$^{-1}$ and 3000 fb$^{-1}$, using the uncertainties reported in Tab.~1 of \cite{Peskin:2013xra} and, as a first step,  assuming  that the central value of the measurements in every channel coincides with the predictions of the SM. In Fig.~\ref{Chifuture} we report the two cases ``CMS-II'' (300 fb$^{-1}$) and ``CMS-HL-II'' (3000 fb$^{-1}$).

\begin{figure}[t]
\centering
\includegraphics[width=1\textwidth]{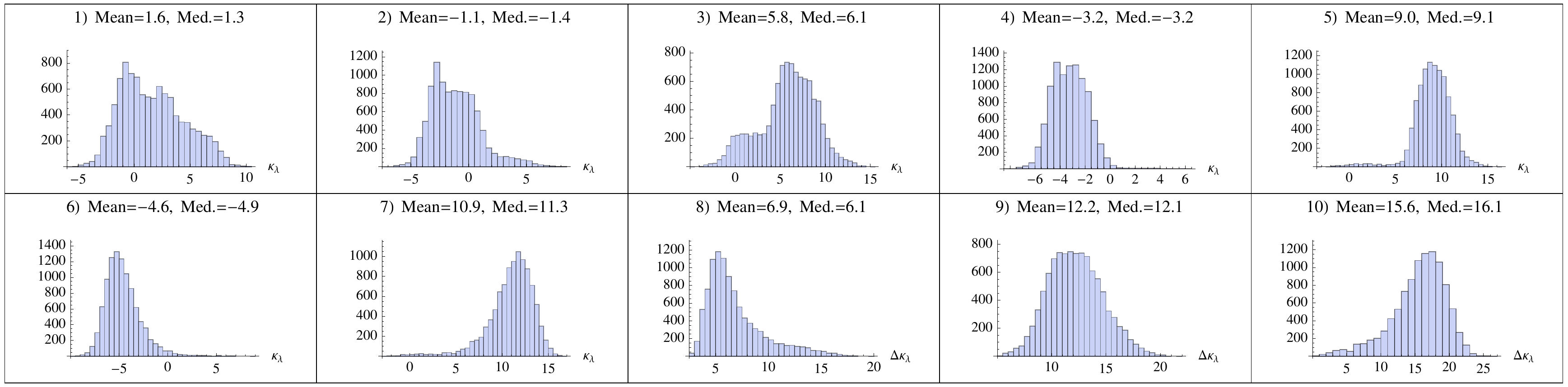}
\caption{Histograms for ``CMS-II'' (300 fb$^{-1}$). The distributions represented are, from left to right and from top to bottom: 1) best values, 2) $1\sigma$  region lower limit, 3) $1\sigma$ region upper limit, 4) $2\sigma$ region lower limit, 5) $2\sigma$ region upper limit, 6) $p>0.05$ region lower limit, 7) $p>0.05$ region upper limit, 8) $1\sigma$  region width, 9) $2\sigma$ region width, 10) $p>0.05$ region width.}
\label{histo300}
\end{figure}

\begin{figure}[t]
\centering
\includegraphics[width=1\textwidth]{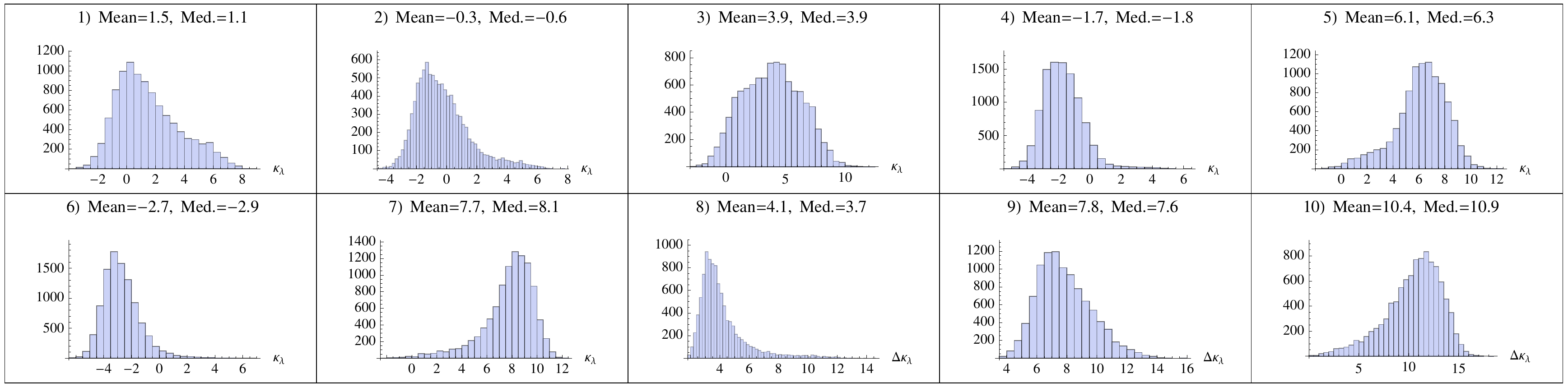}
\caption{As Fig.~\ref{histo300} for  ``CMS-HL-II'' (3000 fb$^{-1}$).}
\label{histo3000}
\end{figure}

Within this approach, best values are by definition: $\ktre^{\rm best}=1$. For the $1\sigma$ and $2\sigma$ intervals, and for the region where the $p$-value is larger than $0.05$, we find that  the ``CMS-II'' (300 fb$^{-1}$) case gives
 \begin{equation} 
\ktre^{1 \sigma} = [-1.8,7.3]\,, ~~~~ \ktre^{2 \sigma} = [-3.5,9.6]\,,~~~~ \ktre^{p>0.05}=[-6.7,13.8]\,,
\label{eq:sigA}
\end{equation}
while for the  ``CMS-HL-II'' (3000 fb$^{-1}$) we obtain 
   \begin{equation} 
\ktre^{1 \sigma} = [-0.7,4.2]\,, ~~~~ \ktre^{2 \sigma} = [-2.0,6.8]\,,~~~~\ktre^{p>0.05}=[-4.1,9.8]\,.
\label{eq:sigB}
\end{equation}
This simplified approach provides a first (rough) idea of the typical intervals that can be expected. A more
reliable approach consists of considering, still within the SM assumption, all the possible central values that could be measured. To this aim,  we produce a collection of pseudo-measurements $\{ \muifexp \}$, where each $\muifexp$ is randomly generated with a gaussian distribution around the SM with a standard deviation equal to the experimental uncertainty cited in Tab.~1 of \cite{Peskin:2013xra}. For each pseudo-experiment we perform a fit and we determine $\ktre^{\rm best}$ and the $\ktre^{1 \sigma}$, $\ktre^{2 \sigma}$  and $\ktre^{p>0.05}$ intervals. In Figs.~\ref{histo300} and \ref{histo3000} we report the results out of a collection of $n=10000$ pseudo-experiment. Frequency histograms together with corresponding mean and median values are  provided for $\ktre^{\rm best}$ and all the extremes and widths of the $\ktre^{1 \sigma}$, $\ktre^{2 \sigma}$  and $\ktre^{p>0.05}$ intervals. From these plots it is clear that most likely the limits written in Eq.~(\ref{eq:sigA}) and (\ref{eq:sigB}) are pessimistic, and the LHC should be able to put even stronger bounds.

\begin{figure}[t]
\centering
\includegraphics[width=0.45\textwidth]{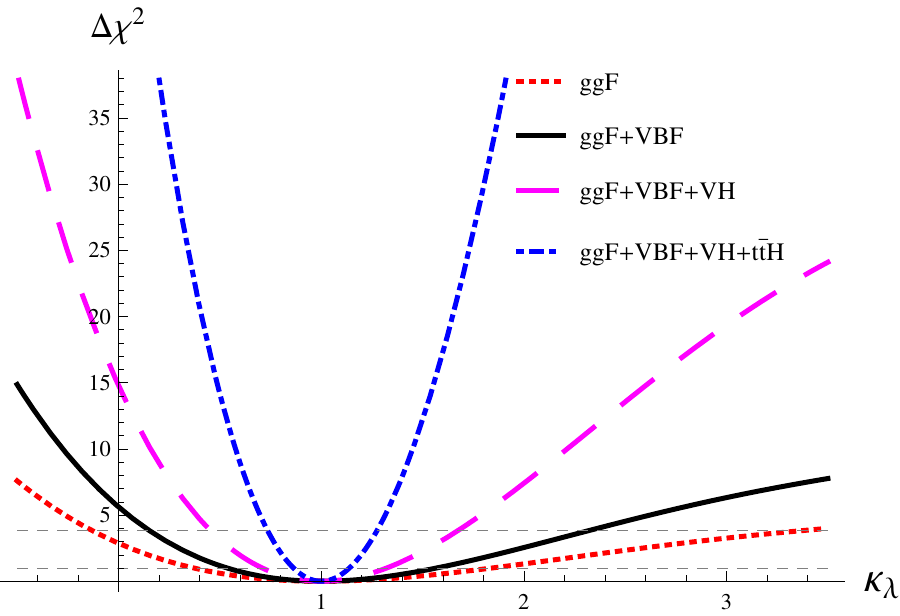}
\includegraphics[width=0.45\textwidth]{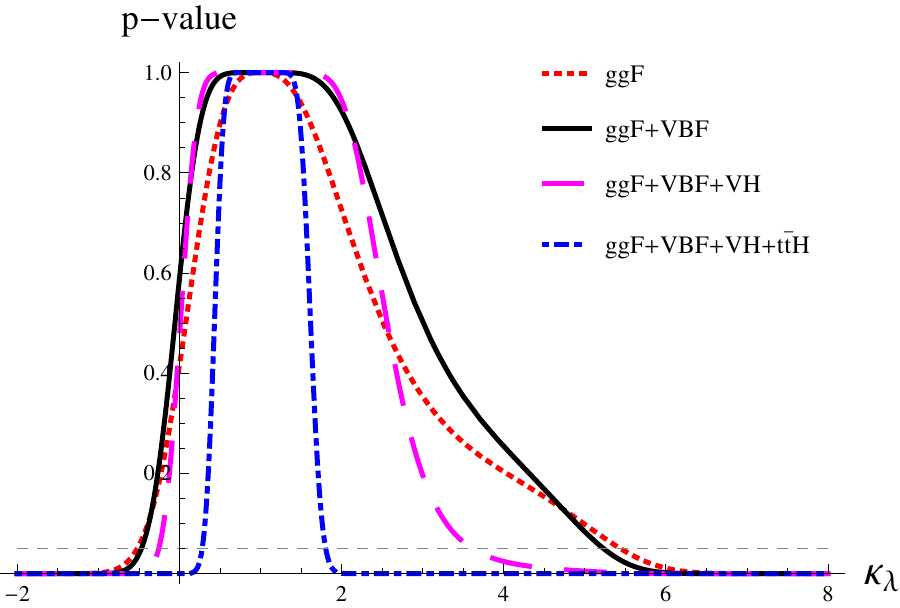}
\caption{In the left and right plots, respectively $\chi^2(\ktre)$ and $p$-value$(\ktre)$ for the P$_{1,2,3,4}$ scenarios with relative uncertainties set at $0.01$.}
\label{Chi1per}
\end{figure}

\begin{figure}[t]
\centering
\includegraphics[width=1\textwidth]{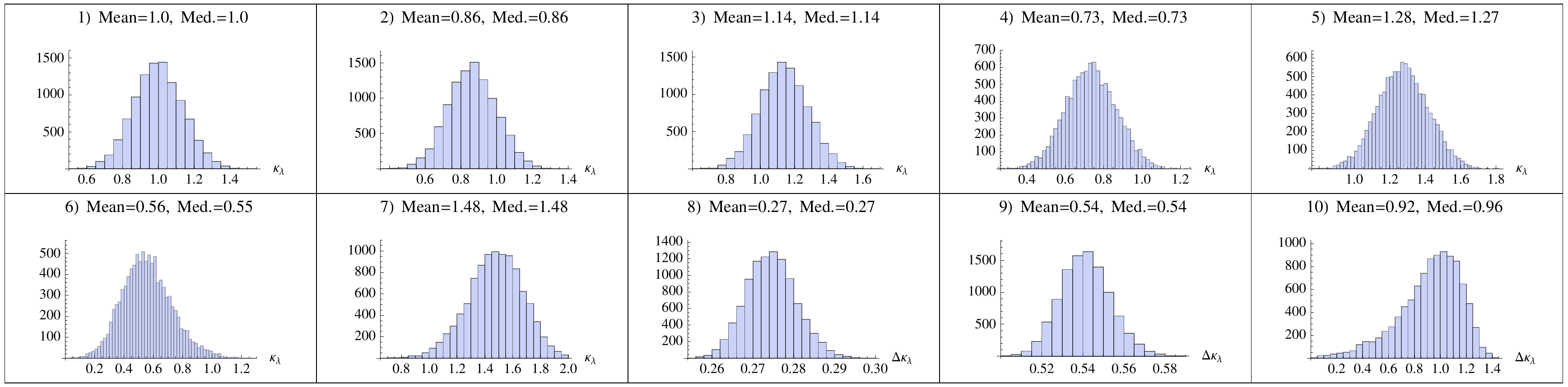}
\caption{As Fig.~\ref{histo300} for the P$_4$ scenario with relative uncertainties set to 0.01.}
\label{histo1per}
\end{figure}

As a last exercise, we consider an optimistic scenario where the quadratic sum of the experimental and theoretical uncertainties amounts to one percent in total. To this aim we employ the observables included in the data sets P$_{1,2,3,4}$, and assume, as first step,  that the measured signal strength is the one of the SM with an associated $0.01$ relative uncertainty. In Fig.~\ref{Chi1per} we report the  obtained $\chi^2(\ktre)$ and $p$-value$(\ktre)$. As expected, a precise measurement of the $\tth$ would lead to a sizeable improvement in the fit. For example, we find that for the scenario P$_4$
 \begin{equation} 
\ktre^{1 \sigma} = [0.86,1.14]\,, ~~~~ \ktre^{2 \sigma} = [0.74,1.28]\,,~~~~ \ktre^{p>0.05}=[0.28,1.80]\,.
\label{eq:fut1}
\end{equation}
Considering as before $n=10000$ pseudo-measurements, the histograms analogous to  those in Fig.~\ref{histo300}
and \ref{histo3000} are shown in Fig.~\ref{histo1per}. Again, we find the indication that, most-likely, in this optimistic scenario stronger bounds  than those reported in Eq.~\eqref{eq:fut1} could be set.

\section{Conclusions}

The structure and properties of the scalar sector encompassing the
observed Higgs boson are largely unexplored and their determination is
one of the major goals of the LHC and future colliders. In the
standard model the Higgs self couplings, trilinear and quartic, are
fixed by the Higgs mass, yet they could be different in scenarios
featuring extended scalar sectors or new strong dynamics. The
most-beaten path to determine the trilinear coupling is via the direct
measurement of Higgs pair production total cross sections and
differential distributions. However, the small expected rates, the
mild dependence of the cross section on the trilinear coupling and the
difficulty of selecting signal from backgrounds make this path very
arduous.

In this work we have put forward an alternative method, which relies
on the effects that loops featuring an anomalous trilinear coupling
would imprint on single Higgs production channels at the LHC. We have
calculated the contributions arising at NLO on all the
phenomenologically relevant single Higgs production ($\ggF$, VBF,
$WH$, $ZH$, $t\bar tH$) and decay ($\gamma \gamma$, $WW^{*}/ZZ^{*}\to
4f$, $b \bar b$, $\tau \tau$) modes at the LHC. Remarkably, we have
found that the $\lambda_3$ dependence is different for each channel
(production times decay) and is also affected by the final state
kinematic configurations. We have then estimated the sensitivity to
the trilinear coupling via a one-parameter fit to the complete set of
single Higgs inclusive measurements at the LHC 8 TeV.  The bounds
obtained are found to be competitive with the current ones obtained
from Higgs pair production. We have also estimated the constraints
that can be obtained at the end of the current Run II and also in the
HL phase with an integrated luminosity of 3000 fb$^{-1}$ expected. In
all cases, the determination of the Higgs self coupling via loop
effects is competitive with the direct determination and will provide
complementary information. 

We remark that when an analysis based on a single observable is made, the
  effects induced by a modification of the trilinear coupling cannot be
  distinguished from those induced by an overall rescaling factor of the
  relevant Higgs coupling, like a $\kappa_f$ or $\kappa_V$ factor. Instead,
  the simultaneous analysis of several observables allows the identification
  of the different sources of the various effects. We also note that, even 
though not exploited in this first study, differential information from single
Higgs production and/or decays could also be used to improve the
sensitivity.

The indirect approach outlined in this work relies on the assumption
that the leading effects from physics beyond the Standard Model affect
the Higgs potential only, {\it i.e.}, the couplings to fermions and
vector bosons are not (or just mildly) affected by new physics at the
tree level.  Admittedly, this might be a limitation for studying some
specific new physics scenarios. However, this assumption is not a requirement for our
  method to be applied.  As information on the Higgs couplings to
  vector boson and the top quark will become more accurate, one could
  think of progressively lift the condition on the other Higgs
  couplings to be SM and allow for tree-level deviations in the global
  fit.  A first straightforward step will be the extension to a
  three-parameter ($\kappa_V,\kappa_f,\kappa_\lambda$) fit, being
  $\kappa_V$, $\kappa_f$ the universal rescaling factors of the
  fermion/boson Higgs couplings. A further step will be the study of
  the additional sensitivity given by the inclusion of collider energy
  and differential observable dependences in the fit.  Work in this
  direction is in progress.
  
 In this work we have chosen to present the results 
  in the context of the  $\kappa$-framework, because with the current sensitivities 
  only rather large deviations from the SM can be probed. Moreover, in this way our results can be 
  straightforwardly implemented in the experimental global analyses~\cite{Khachatryan:2016vau}, 
  which are also currently based on the $\kappa$-framework. The next step will be the
  interpretation of our loop calculations in the context of an effective field theory including at least dimension-6 operators. In this context, issues such as how many independent observables are needed to lift all possible degeneracies in the effects induced by different operators (at tree- and one-loop level), need further investigation.

\section*{Acknowledgements}
We would like to thank the LHCXSWG for providing a stimulating environment. 
P.P.G. would like to thank Sally Dawson for useful comments and discussions.   D.P. and F.M. are thankful to Giacomo Bruno for many patient explanations. This work is supported in part (D.P. and F.M.)  by the ERC grant 291377 ``LHCtheory: Theoretical predictions and analyses of LHC physics: advancing the precision frontier", by the IISN ``Fundamental interactions'' convention 4.4517.08, and by the Belgian Federal Science Policy Office through the Interuniversity Attraction Pole P7/37. The work of P.P.G. was supported by the United States Department of Energy under Grant Contracts de-sc0012704.

\clearpage

\appendix

\begin{appendletterA}
\section{Comparison with the EFT approach}
\label{sec:EFT}
The SM potential for the Higgs doublet field reads
\begin{equation}\label{VSM}
V^{\rm SM}(\Phi)=-\mu^2 (\Phi^\dagger \Phi)+\lambda (\Phi^\dagger \Phi)^2\, ,\qquad \Phi = \frac{1}{\sqrt{2}} \begin{pmatrix}\phi^+ \cr    v+H+i \phi^0\, 
 \end{pmatrix}\  ,
\end{equation}
and can be modified by adding the dimension-6 operators $(\Phi^\dagger \Phi)^3$,
\begin{equation}\label{Vdim6}
V^{\rm dim-6}(\Phi)=V^{\rm SM}(\Phi) + \frac{c_6}{v^2}(\Phi^\dagger \Phi)^3\, ,
\end{equation}
where the normalization of the operator $(\Phi^\dagger \Phi)^3$ is
$v=(\sqrt2 G_\mu)^{-1/2}=246~\gev$. The relations among $\mh$, $v$,
$\mu$ and $\lambda$ are different in $V^{\rm SM}(\Phi)$ and $V^{\rm
  dim-6}(\Phi)$. We determine $\lambda$ and $\mu$ as function of the
measured quantities, $\mh$ and $v$, and of the new parameter
$c_{6}$. Once all the dependences are expressed as function of $\mh$,
$v$ and $c_6$, we can derive the value of the coefficient in front of
$H^3$ which in the paper is called $\tril$, as well as the coefficient
in front of the quartic term $H^4$, which is denoted as $\qual$. The
SM relations are recovered by setting $c_6=0$.

With the condition 
$\frac{dV^{\rm dim-6}(\Phi)}{d\Phi} \Big |_{|\Phi| = v/ \sqrt{2}}=0$ , one obtains
\begin{equation}\label{vasmul}
v=\frac{2\mu}{\sqrt{4\lambda +3 c_6 }}~\TO~ \mu=\frac{1}{2}v\sqrt{4\lambda+3c_6}\, ,
\end{equation}
which  after Electroweak Symmetry Breaking implies
\begin{equation}\label{mhasvl}
\mh^2=v^2 (2\lambda + 3 c_6)\, \TO \lambda=\frac{\mh^2}{2 v^2}  -\frac{3c_6}{2}\, ,
\end{equation}
and
\begin{equation}\label{trildadim6}
c_{H^3}\equiv v \tril=v\left(\lambda+\frac{5}{2} c_6\right)=\frac{\mh^2}{2v}+c_6 v \TO\ktre=1+\frac{2 c_6 v^2}{\mh^2}\,.
\end{equation}
At a first sight, the linear relation in Eq.~\eqref{trildadim6} seems
to imply that with the potential $V^{\rm dim-6}(\Phi)$ any value of
$\tril$ can be obtained. However, one can require that the
potential is bounded from below\footnote{Here we are not taking into
  account Renormalization-Group-Equation (RGE) effects on $\lambda$
  and $c_6$, which may add additional constraints; only the potential
  without quantum effects is considered.} ($c_6 > 0$) and that $v$ is
the global minimum. The latter condition had been
already discussed in Ref.~\cite{Grojean:2004xa} and can be easily
derived substituting in the potential of Eq.~\eqref{Vdim6} $\mu$ and
$\lambda$ with $\mh$ and $v$ via Eqs.~\eqref{vasmul} and
\eqref{mhasvl}:
\begin{equation}\label{Vdim6_v2}
V^{\rm dim-6}(\Phi)=\left(-\frac{\mh^2 }{2}  + 
\frac{3}{4} c_6 v^2 \right)\Phi^\dagger \Phi\ + \left(\frac{\mh^2 }{2 v^2}  - 
\frac{3}{2} c_6  \right)(\Phi^\dagger \Phi)^2 + 
\frac{c_6}{v^2}(\Phi^\dagger \Phi)^3\, .
\end{equation}
Since  $\Phi=0$ can be a local minimum, the condition that $v$ is a global
minimum requires
\begin{equation}
V^{\rm dim-6}(v/\sqrt{2})=\frac{c_6 v^4 - \mh^2 v^2}{8} <0=V^{\rm dim-6}(0)\, .
\end{equation}
or $c_6<\mh^2/v^2$. 
Thus, with the inclusion of only the $(\Phi^\dagger \Phi)^3$ operator in the SM 
Lagrangian $\ktre$ is constrained to be in  the range
\begin{equation}
1<\ktre<3 \label{klinc6}\,.
\end{equation}
It is worth to notice that this bound has been derived without any
assumption on the size $c_6$, which in an EFT approach would be
subject to further constraints depending on the scale of new physics
$\Lambda$.

In a general EFT approach in principle the value of $\tril$ can be
affected also by another dimension-6 operator, namely,
$\frac{c_\Phi}{2v^2}\partial^\mu(\Phi^\dagger \Phi)
\partial_\mu(\Phi^\dagger \Phi)$. However, other couplings of the
Higgs boson would also be affected by this operator, such as the
coupling with the $Z$ boson and with the fermions. Thus, these effects
would be already present at LO in single-Higgs production and would be
in general much larger than the effects induced by an anomalous
$\tril$ coupling. Only for values $1<\ktre<3$ and assuming $c_\Phi=0$
the results obtained in this paper can be converted to values of $c_6$
via eq.~\eqref{trildadim6}.  Moreover, in the EFT approach, Wilson
coefficients at the scale $\Lambda$ are typically expected to be smaller in absolute value
than $4\pi$. This requirement would additionally set the
constraint
\begin{equation}
c_6 < 4\pi \frac{v^2}{\Lambda^2} \TO 1<\ktre< \min \left( 3\,, 1+ 8\pi \frac{v^4}{\mh^2 \Lambda^2} \right)\,.\label{klinc6Lambda}
\end{equation}

Analogously to what has been done for the trilinear coupling, we can
define $\qual\equiv\kqual \qual^{\rm SM}$ finding
\begin{equation}
\kqual= 1 + \frac{12 c_6 v^2}{\mh^2}\,,  \label{kl4fromc6}
\end{equation}
which implies
\begin{equation}
 \kqual= 6 \ktre - 5  \TO 1 < \kqual < \min \left( 13\,, 1+ 48\pi \frac{v^4}{\mh^2 \Lambda^2} \right)\, \label{kltokl4}\,,
\end{equation}
since, with the $V^{\rm dim-6}(\Phi)$ potential, $\qual$ is a prediction fixed by $\mh,v$ and $\tril$. 

As last comments concerning the potential in Eq.~\eqref{Vdim6}, we
want to stress that the constraints in
Eqs.~\eqref{klinc6}-\eqref{klinc6Lambda}, the relation between $\tril$
and $\qual$ and thus also the constraints on $\qual$ in
Eq.~\eqref{kltokl4} are parametrisation independent, {\it i.e.}, they
are not altered by the choice of normalisation of the $(\Phi^\dagger
\Phi)^3$ operator. Using for instance, the normalisation $\bar{c}_6
\frac{\lambda}{v^2}$ of Ref.~\cite{Gorbahn:2016uoy},
Eqs.~\eqref{vasmul}-\eqref{trildadim6} and Eq.~\eqref{kl4fromc6} would
change, namely:
\begin{equation}\label{mhasvl_old}
\mh^2=v^2 \lambda (2 + 3 c_6)\TO \lambda=\frac{\mh^2}{v^2(2+3 \bar{c}_6)}\,
\end{equation}
\begin{equation}\label{klold}
\ktre=\frac{2+5 \bar{c}_6 }{2+3 \bar{c}_6 }\, ,
\end{equation}
\begin{equation}\label{kl4old}
 \kqual=\frac{2+15 \bar{c}_6 }{2+3 \bar{c}_6 }\, .
\end{equation}
Equations \eqref{klold} and \eqref{kl4old} can be easily related to
\eqref{trildadim6} and \eqref{kl4fromc6} in the limit $c_6$ or
$\bar{c}_6\TO 0$, {\it i.e.}, $\ktre,~\kqual \sim 1$. On the other
hand, with this parametrisation, it is less obvious how to determine
the maximal and minimal possible values for $\ktre$. In any case,
imposing the conditions that the potential is bounded from below and
that $v$ is the global minimum, it is possible to recover the
bound $1<\ktre<3$, confirming its independence on the choice of
normalisation of the $(\Phi^\dagger \Phi)^3$ term.

\medskip 

As a final exercise, we consider the extension of the SM potential $V^{\rm SM}$
\begin{equation}\label{Vdim8}
V^{\rm dim-8}(\Phi)=V^{\rm SM}(\Phi) + \frac{c_6}{v^2}(\Phi^\dagger \Phi)^3 + \frac{c_8}{v^4}(\Phi^\dagger \Phi)^4\, ,
\end{equation}
where besides the $(\Phi^\dagger \Phi)^3$ term also the $(\Phi^\dagger \Phi)^4$ is included. Relations corresponding to those in Eqs.~\eqref{vasmul}-\eqref{trildadim6} and \eqref{kl4fromc6} can be derived in a completely analogous way. We write them directly as function of $\mh, \lambda, c_6$ and $c_8$, where by setting $c_8=0$ one recovers the analogous ones for the potential in Eq.~\ref{Vdim6}:  
\begin{equation}\label{vasmul-dim8}
\mu^2=\frac{\mh^2}{2} -\frac{3c_6}{4}v^2-c_8 v^2 \, ,
\end{equation}
\begin{equation}\label{mhasvlb--dim8}
\lambda=\frac{\mh^2}{2 v^2}  -\frac{3c_6}{2}-\frac{3c_8}{2}\, ,
\end{equation}
\begin{equation}\label{trildadim8}
\ktre=1+\frac{(2 c_6 + 4 c_8 ) v^2}{\mh^2}\, ,
\end{equation}
\begin{equation}\label{kl4fromc6c8}
\kqual= 1 + \frac{(12 c_6 + 32 c_8) v^2}{\mh^2}\, .  
\end{equation}

At variance with the case of $V^{\rm dim-6}(\Phi)$, with the inclusion of the $\frac{c_8}{v^4}(\Phi^\dagger \Phi)^4$ term the quantity $\kqual$ is independent of $\ktre$, {\it i.e.}, $c_6$ and $c_8$ can be traded off with $\ktre$ and $\kqual$. The requirement that the potential is bounded from below implies $c_8>0$, which in conjunction with the requirement that the global minimum is located at $\Phi=v/\sqrt{2}$ implies
\begin{equation}\label{kqualfromktre-8}
-4+4\ktre+\ktre^2<\kqual <  \frac{-31+30\ktre+9\ktre^2}{8}\, .
\end{equation}
Thus, without any constraint on the size of $c_6$ and $c_8$, such as those coming from an EFT, $\ktre$ is not bounded and $\kqual$ is constrained by Eq.~(\ref{kqualfromktre-8}).

 \end{appendletterA}
\clearpage

\begin{appendletterB}

\section{$C_1$ terms for $\sigma(gg\rightarrow H)$ and 
$\Gamma(H\rightarrow\gamma\gamma)$}
\label{app1}
In this appendix we present the results for the $C_1$ factor 
in the gluon-gluon-fusion Higgs production and in the Higgs 
partial decay into two photons.

\subsection{$\sigma(gg\rightarrow H)$}
We write the SM gluon-gluon-fusion Higgs production partonic cross-section  as
\beqn
\sigma=\frac{G_\mu \alpha_s^2}{512\sqrt{2}\pi}|{\cal G}|^2,
\eeqn
where ${\cal G}={\cal G}^{1l}+{\cal G}^{2l}+...$ with the lowest order 
contribution given by\footnote{The analytic continuation is obtained
with the replacement $-\mh^2 \to -\mh^2 - i \epsilon$}
\beq
{\cal G}^{1l} = -\frac{4}{h_t}\left(2- \frac{1-4/h_t}2
\log^2\left[\frac{\sqrt{1- 4/h_t}-1}{\sqrt{1- 4/h_t}+1}\right]
\right),~~~~h_t \equiv \frac{\mh^2}{\mt^2}~.
\label {G1l}
\eeq 

The two-loop contribution can be written as: ${\cal G}^{2l}= K_r\,{\cal G}^{1l}
+ {\cal G}^{2l}_{1{\rm PI}}$ with $ {\cal G}^{2l}_{1{\rm PI}}$ the contribution of the
one-particle irreducible (1PI) vertex diagrams and 
\beq
K_r \equiv \left[ \frac{A_{\sss WW}}{\mw^2} - V - B + (\delta Z_H)_{\sss {\rm SM}} \right]\,,
\eeq
where $A_{\sss WW}$ is the transverse part of the $W$ self-energy at zero 
momentum transverse, the quantities $V$ and $B$ represent the vertex and 
box corrections in the $\mu$-decay amplitude and  $ (\delta Z_H)_{\sss {\rm SM}}$ is
the Higgs field wave function renormalisation constant in the SM. 

In our scenario the modification of the Higgs wave function, represented
by the $C_2$ coefficient, will affect the $K_r$ term while $C_1$ is extracted 
from the diagrams in Fig.~\ref{fig3} that contribute to $ {\cal G}^{2l}_{1{\rm PI}}$.  

Under the standard approximation  of the factorisation of the EW corrections in
$\sigma (gg\rightarrow H)$ we have for $C_1$ 
\beq
C_1^{\sigma}(\ggF)= 
2 \frac{  {\cal G}^{2l}_{1{\rm PI}, \,{\trilsm}}}{{\cal G}^{1l}}\,,
\label{C1ggh}
\eeq
where
\beqn
{\cal G}^{2l}_{1{\rm PI}, \,{\trilsm}}&=&\frac{G_\mu \mh^2}{2\, \sqrt{2} \pi^2}
\left[ \frac{-23 + 4\sqrt{3}\pi}{24} + \frac12 \log(h_t)  
\right. \nonumber\\
&&  ~~~~~+ h_t \left( \frac{7}{480}(-37 + 4\sqrt{3}\pi) + 
\frac{7}{20}\log(h_t)\right)
\nonumber\\
 &&  ~~~~~+ h_t^2 \left( \frac{-464419 + 33810\sqrt{3}\pi}{2116800}
+ \frac{349}{2016}\log(h_t)\right)\nonumber\\ 
 &&  ~~~~+ \left . h_t^3\left(-\frac{31795373}{381024000} + 
\frac{13 \pi}{1050\sqrt{3}} +   
\frac{1741}{21600} \log(h_t) \right) \right]~.
\eeqn

\subsection{$\Gamma(H\rightarrow\gamma\gamma)$}

For $\Gamma(H\rightarrow\gamma\gamma)$ we have
\beq
\Gamma=\frac{G_\mu \alpha^2 M_h^3}{128\sqrt{2}\pi^3}|F|^2,
\eeq
with
${\cal F}={\cal F}^{1l}+{\cal F}^{2l}+...$. The lowest order 
contribution is given by
\beq
{\cal F}^{1l}= N_c Q^2 {\cal G}^{1l} +
 2(1+\frac6{\hw})-\frac6{\hw}(1-\frac2{\hw}) 
\log^2\left[\frac{\sqrt{1- 4/\hw}-1}{\sqrt{1- 4/\hw}+1}\right]\,,
\eeq
with $Q=2/3,\: N_c =3 $ and $\hw= \mh^2/\mw^2$.

The two-loop form factor ${\cal F}^{2l}$ can be decomposed in the same
way as ${\cal G}^{2l}$ so that $C_1$ can be extracted from the 1PI diagrams 
in Figs.\ref{fig3} and \ref{fig4})  evaluated in the unitary gauge. We find
\beq
C_1^{\Gamma}(\gamma\gamma)= 
2 \frac{  {\cal F}^{2l}_{1{\rm PI}, \, \trilsm}}{{\cal F}^{1l}}\,.
\label{C1Gaa}
\eeq
where 
\beqn
{\cal F}^{2l}_{1{\rm PI}, \, \trilsm} &=& 
N_c Q^2 {\cal G}^{2l}_{\trilsm} + \frac{G_\mu \mw^2}{2\, \sqrt{2} \pi^2}\Bigg\{p_w^2 \bigg [-36 + 12 \hw - 15 \hw^2 + \frac{9}{2} \hw^3   \nonumber\\
&-& 12(6 - 46 \hw + 13 \hw^2)  L_w  + 9 (-8 - 12 \hw - 6 \hw^2 + 3 \hw^3) \phi_w \bigg ] \nonumber \\
&+& p_w^4 \bigg [\frac{1}{30} (-38880 + 98640 \hw - 68384 \hw^2 + 15204 \hw^3 + 142 \hw^4 - 308 \hw^5 + 33 \hw^6) \nonumber\\
  &-&  \frac{2}{15} (19440 - 26760 \hw + 15028 \hw^2 - 7262 \hw^3 + 1522 \hw^4 + 57 \hw^5)L_w  \nonumber\\
  &+&  8 (-324 + 500 \hw - 323 \hw^2 + 102 \hw^3 - 31 \hw^4 + 7 \hw^5) \phi_w\bigg ]\nonumber\\
  &+& p_w^6\bigg [\frac{1}{945} (-38283840 + 84825216 \hw - 70055664 \hw^2 + 18977592 \hw^3 - 2081216 \hw^4 \nonumber\\
  &+& 252530 \hw^5 - 56436 \hw^6 + 54710 \hw^7 - 9158 \hw^8 + 513 \hw^9)\nonumber\\
   &-& \frac{2}{105}(4253760 - 9166080 \hw + 8167712 \hw^2 - 5453632 \hw^3\nonumber\\
     &+&  1553124 \hw^4 - 298912 \hw^5 + 78152 \hw^6 - 3992 \hw^7 + 171 \hw^8) L_w  \nonumber\\
   &+& \frac{8}{3} (-30384 + 70536 \hw - 69084 \hw^2 + 34642 \hw^3\nonumber\\
    &-& 13138 \hw^4 + 2337 \hw^5 - 82 \hw^6 + 43 \hw^7)\phi_w \bigg ] \nonumber\\
    &+& p_w^8 \bigg [\frac{1}{4725}(-6078844800 + 15433978560 \hw \nonumber\\
    &-& 16158069376 \hw^2 +  9535767472 \hw^3  \nonumber \\
    &-& 3860103960 \hw^4 + 933792696 \hw^5 - 198236360 \hw^6 + 49562148 \hw^7 \nonumber \\
    &+& 370584 \hw^8 - 1829312 \hw^9 + 410373 \hw^{10} - 40412 \hw^{11} + 1566 \hw^{12}) \nonumber \\
    &-& \frac{4}{1575} (1013140800 - 2714896800 \hw  \nonumber \\
     &+& 3103464560 \hw^2 - 1987417480 \hw^3 + 754138872 \hw^4  \nonumber \\
    &-& 219727216 \hw^5 + 5585768 \hw^6 + 15961770 \hw^7 \nonumber \\
    &-& 1982560 \hw^8 + 349052 \hw^9 - 25056 \hw^{10} + 783 \hw^{11}) L_w \nonumber \\
    &+& \frac{32}{15}(-1206120 + 3433040 \hw - 4226570 \hw^2 + 2964582 \hw^3 - 1314797 \hw^4  \nonumber \\
    &+& 372126 \hw^5 - 99064 \hw^6 + 16782 \hw^7 + 662 \hw^8 + 121 \hw^9)\phi_w  \bigg ] \Bigg\}\nonumber\,,
\eeqn
where
$p_w^2=\frac{q^2}{4\mw^2}\frac{1}{\hw(\hw-4)^2}$,
with $q^2$   the squared external momentum of the Higgs field that is put
on the mass-shell at the end of the calculation, $q^2= \mh^2$, and
$L_w=\frac{\log(h_w)}{(\hw-4)}$,
$\phi_w=\phi(\frac{h_w}{4})\frac{1}{\hw(\hw-4)}$,
 with
\beq
\phi(z)=
      4\sqrt{\frac{z}{1-z}}\ \text{Im}(\text{Li}_2 (e^{i 2\arcsin(\sqrt{z})}))~. 
\eeq

 \end{appendletterB}

\bibliographystyle{JHEP}
\bibliography{DGMP}

\providecommand{\href}[2]{#2}\begingroup\raggedright\begin{thebibliography}{10}

\bibitem{Chatrchyan:2012xdj}
{\scshape CMS} collaboration, S.~Chatrchyan et~al., \emph{{Observation of a new
  boson at a mass of 125 GeV with the CMS experiment at the LHC}},
  \href{http://dx.doi.org/10.1016/j.physletb.2012.08.021}{\emph{Phys. Lett.}
  {\bf B716} (2012) 30--61}, [\href{http://arxiv.org/abs/1207.7235}{{\tt
  1207.7235}}].

\bibitem{Aad:2012tfa}
{\scshape ATLAS} collaboration, G.~Aad et~al., \emph{{Observation
  of a new particle in the search for the Standard Model Higgs boson with the
  ATLAS detector at the LHC}},
  \href{http://dx.doi.org/10.1016/j.physletb.2012.08.020}{\emph{Phys.Lett.}
  {\bf B716} (2012) 1--29}, [\href{http://arxiv.org/abs/1207.7214}{{\tt
  1207.7214}}].

\bibitem{Khachatryan:2014jba}
{\scshape CMS} collaboration, V.~Khachatryan et~al., \emph{{Precise
  determination of the mass of the Higgs boson and tests of compatibility of
  its couplings with the standard model predictions using proton collisions at
  7 and 8 $\,\text {TeV}$}},
  \href{http://dx.doi.org/10.1140/epjc/s10052-015-3351-7}{\emph{Eur. Phys. J.}
  {\bf C75} (2015) 212}, [\href{http://arxiv.org/abs/1412.8662}{{\tt
  1412.8662}}].

\bibitem{Aad:2015gba}
{\scshape ATLAS} collaboration, G.~Aad et~al., \emph{{Measurements of the Higgs
  boson production and decay rates and coupling strengths using pp collision
  data at $\sqrt{s}=7$ and 8 TeV in the ATLAS experiment}},
  \href{http://dx.doi.org/10.1140/epjc/s10052-015-3769-y}{\emph{Eur. Phys. J.}
  {\bf C76} (2016) 6}, [\href{http://arxiv.org/abs/1507.04548}{{\tt
  1507.04548}}].

\bibitem{Khachatryan:2016vau}
{\scshape ATLAS, CMS} collaboration, G.~Aad et~al., \emph{{Measurements of the
  Higgs boson production and decay rates and constraints on its couplings from
  a combined ATLAS and CMS analysis of the LHC $pp$ collision data at
  $\sqrt{s}=$ 7 and 8 TeV}},  \href{http://arxiv.org/abs/1606.02266}{{\tt
  1606.02266}}.

\bibitem{LHCHiggsCrossSectionWorkingGroup:2012nn}
{\scshape LHC Higgs Cross Section Working Group} collaboration, A.~David,
  A.~Denner, M.~Duehrssen, M.~Grazzini, C.~Grojean, G.~Passarino et~al.,
  \emph{{LHC HXSWG interim recommendations to explore the coupling structure of
  a Higgs-like particle}},  \href{http://arxiv.org/abs/1209.0040}{{\tt
  1209.0040}}.

\bibitem{Heinemeyer:2013tqa}
{\scshape LHC Higgs Cross Section Working Group} collaboration, J.~R. Andersen
  et~al., \emph{{Handbook of LHC Higgs Cross Sections: 3. Higgs Properties}},
  \href{http://arxiv.org/abs/1307.1347}{{\tt 1307.1347}}.

\bibitem{CMS:2013xfa}
{\scshape CMS} collaboration, \emph{{Projected Performance of an Upgraded CMS
  Detector at the LHC and HL-LHC: Contribution to the Snowmass Process}},  in
  \emph{{Community Summer Study 2013: Snowmass on the Mississippi (CSS2013)
  Minneapolis, MN, USA, July 29-August 6, 2013}}, 2013.
\newblock \href{http://arxiv.org/abs/1307.7135}{{\tt 1307.7135}}.

\bibitem{Peskin:2013xra}
M.~E. Peskin, \emph{{Estimation of LHC and ILC Capabilities for Precision Higgs
  Boson Coupling Measurements}},  in \emph{{Community Summer Study 2013:
  Snowmass on the Mississippi (CSS2013) Minneapolis, MN, USA, July 29-August 6,
  2013}}, 2013.
\newblock \href{http://arxiv.org/abs/1312.4974}{{\tt 1312.4974}}.

\bibitem{Gupta:2013zza}
R.~S. Gupta, H.~Rzehak and J.~D. Wells, \emph{{How well do we need to measure
  the Higgs boson mass and self-coupling?}},
  \href{http://dx.doi.org/10.1103/PhysRevD.88.055024}{\emph{Phys. Rev.} {\bf
  D88} (2013) 055024}, [\href{http://arxiv.org/abs/1305.6397}{{\tt
  1305.6397}}].

\bibitem{Efrati:2014uta}
A.~Efrati and Y.~Nir, \emph{{What if $\lambda_{hhh}\neq 3m_h^2/v$}},
  \href{http://arxiv.org/abs/1401.0935}{{\tt 1401.0935}}.

\bibitem{Anastasiou:2016cez}
C.~Anastasiou, C.~Duhr, F.~Dulat, E.~Furlan, T.~Gehrmann, F.~Herzog et~al.,
  \emph{{High precision determination of the gluon fusion Higgs boson
  cross-section at the LHC}},  \href{http://arxiv.org/abs/1602.00695}{{\tt
  1602.00695}}.

\bibitem{deFlorian:2013jea}
D.~de~Florian and J.~Mazzitelli, \emph{{Higgs Boson Pair Production at
  Next-to-Next-to-Leading Order in QCD}},
  \href{http://dx.doi.org/10.1103/PhysRevLett.111.201801}{\emph{Phys. Rev.
  Lett.} {\bf 111} (2013) 201801}, [\href{http://arxiv.org/abs/1309.6594}{{\tt
  1309.6594}}].

\bibitem{Maltoni:2014eza}
F.~Maltoni, E.~Vryonidou and M.~Zaro, \emph{{Top-quark mass effects in double
  and triple Higgs production in gluon-gluon fusion at NLO}},
  \href{http://dx.doi.org/10.1007/JHEP11(2014)079}{\emph{JHEP} {\bf 11} (2014)
  079}, [\href{http://arxiv.org/abs/1408.6542}{{\tt 1408.6542}}].

\bibitem{Borowka:2016ehy}
S.~Borowka, N.~Greiner, G.~Heinrich, S.~Jones, M.~Kerner, J.~Schlenk et~al.,
  \emph{{Higgs Boson Pair Production in Gluon Fusion at Next-to-Leading Order
  with Full Top-Quark Mass Dependence}},
  \href{http://dx.doi.org/10.1103/PhysRevLett.117.012001}{\emph{Phys. Rev.
  Lett.} {\bf 117} (2016) 012001}, [\href{http://arxiv.org/abs/1604.06447}{{\tt
  1604.06447}}].

\bibitem{Baglio:2012np}
J.~Baglio, A.~Djouadi, R.~Gr\"ober, M.~M. M\"uhlleitner, J.~Quevillon and M.~Spira,
  \emph{{The measurement of the Higgs self-coupling at the LHC: theoretical
  status}}, \href{http://dx.doi.org/10.1007/JHEP04(2013)151}{\emph{JHEP} {\bf
  04} (2013) 151}, [\href{http://arxiv.org/abs/1212.5581}{{\tt 1212.5581}}].

\bibitem{Frederix:2014hta}
R.~Frederix, S.~Frixione, V.~Hirschi, F.~Maltoni, O.~Mattelaer, P.~Torrielli
  et~al., \emph{{Higgs pair production at the LHC with NLO and parton-shower
  effects}},
  \href{http://dx.doi.org/10.1016/j.physletb.2014.03.026}{\emph{Phys. Lett.}
  {\bf B732} (2014) 142--149}, [\href{http://arxiv.org/abs/1401.7340}{{\tt
  1401.7340}}].

\bibitem{Baur:2003gp}
U.~Baur, T.~Plehn and D.~L. Rainwater, \emph{{Probing the Higgs selfcoupling at
  hadron colliders using rare decays}},
  \href{http://dx.doi.org/10.1103/PhysRevD.69.053004}{\emph{Phys. Rev.} {\bf
  D69} (2004) 053004}, [\href{http://arxiv.org/abs/hep-ph/0310056}{{\tt
  hep-ph/0310056}}].

\bibitem{Yao:2013ika}
W.~Yao, \emph{{Studies of measuring Higgs self-coupling with $HH\rightarrow
  b\bar b \gamma\gamma$ at the future hadron colliders}},  in \emph{{Community
  Summer Study 2013: Snowmass on the Mississippi (CSS2013) Minneapolis, MN,
  USA, July 29-August 6, 2013}}, 2013.
\newblock \href{http://arxiv.org/abs/1308.6302}{{\tt 1308.6302}}.

\bibitem{Barger:2013jfa}
V.~Barger, L.~L. Everett, C.~B. Jackson and G.~Shaughnessy, \emph{{Higgs-Pair
  Production and Measurement of the Triscalar Coupling at LHC(8,14)}},
  \href{http://dx.doi.org/10.1016/j.physletb.2013.12.013}{\emph{Phys. Lett.}
  {\bf B728} (2014) 433--436}, [\href{http://arxiv.org/abs/1311.2931}{{\tt
  1311.2931}}].

\bibitem{Azatov:2015oxa}
A.~Azatov, R.~Contino, G.~Panico and M.~Son, \emph{{Effective field theory
  analysis of double Higgs boson production via gluon fusion}},
  \href{http://dx.doi.org/10.1103/PhysRevD.92.035001}{\emph{Phys. Rev.} {\bf
  D92} (2015) 035001}, [\href{http://arxiv.org/abs/1502.00539}{{\tt
  1502.00539}}].

\bibitem{Lu:2015jza}
C.-T. Lu, J.~Chang, K.~Cheung and J.~S. Lee, \emph{{An exploratory study of
  Higgs-boson pair production}},
  \href{http://dx.doi.org/10.1007/JHEP08(2015)133}{\emph{JHEP} {\bf 08} (2015)
  133}, [\href{http://arxiv.org/abs/1505.00957}{{\tt 1505.00957}}].

\bibitem{Dolan:2012rv}
M.~J. Dolan, C.~Englert and M.~Spannowsky, \emph{{Higgs self-coupling
  measurements at the LHC}},
  \href{http://dx.doi.org/10.1007/JHEP10(2012)112}{\emph{JHEP} {\bf 10} (2012)
  112}, [\href{http://arxiv.org/abs/1206.5001}{{\tt 1206.5001}}].

\bibitem{Papaefstathiou:2012qe}
A.~Papaefstathiou, L.~L. Yang and J.~Zurita, \emph{{Higgs boson pair production
  at the LHC in the $b \bar{b} W^+ W^-$ channel}},
  \href{http://dx.doi.org/10.1103/PhysRevD.87.011301}{\emph{Phys. Rev.} {\bf
  D87} (2013) 011301}, [\href{http://arxiv.org/abs/1209.1489}{{\tt
  1209.1489}}].

\bibitem{deLima:2014dta}
D.~E. Ferreira~de Lima, A.~Papaefstathiou and M.~Spannowsky, \emph{{Standard
  model Higgs boson pair production in the ( $ b\overline{b} $ )( $
  b\overline{b} $ ) final state}},
  \href{http://dx.doi.org/10.1007/JHEP08(2014)030}{\emph{JHEP} {\bf 08} (2014)
  030}, [\href{http://arxiv.org/abs/1404.7139}{{\tt 1404.7139}}].

\bibitem{Wardrope:2014kya}
D.~Wardrope, E.~Jansen, N.~Konstantinidis, B.~Cooper, R.~Falla and
  N.~Norjoharuddeen, \emph{{Non-resonant Higgs-pair production in the
  $b\overline{b}$ $b\overline{b}$ final state at the LHC}},
  \href{http://dx.doi.org/10.1140/epjc/s10052-015-3439-0}{\emph{Eur. Phys. J.}
  {\bf C75} (2015) 219}, [\href{http://arxiv.org/abs/1410.2794}{{\tt
  1410.2794}}].

\bibitem{Behr:2015oqq}
J.~K. Behr, D.~Bortoletto, J.~A. Frost, N.~P. Hartland, C.~Issever and J.~Rojo,
  \emph{{Boosting Higgs pair production in the $b\bar{b}b\bar{b}$ final state
  with multivariate techniques}},  \href{http://arxiv.org/abs/1512.08928}{{\tt
  1512.08928}}.

\bibitem{Englert:2014uqa}
C.~Englert, F.~Krauss, M.~Spannowsky and J.~Thompson, \emph{{Di-Higgs
  phenomenology in $t\bar{t}hh$: The forgotten channel}},
  \href{http://dx.doi.org/10.1016/j.physletb.2015.02.041}{\emph{Phys. Lett.}
  {\bf B743} (2015) 93--97}, [\href{http://arxiv.org/abs/1409.8074}{{\tt
  1409.8074}}].

\bibitem{Liu:2014rva}
T.~Liu and H.~Zhang, \emph{{Measuring Di-Higgs Physics via the $t \bar t hh \to
  t \bar t b \bar bb\bar b$ Channel}},
  \href{http://arxiv.org/abs/1410.1855}{{\tt 1410.1855}}.

\bibitem{Cao:2015oxx}
Q.-H. Cao, Y.~Liu and B.~Yan, \emph{{Measuring Trilinear Higgs Coupling in
  $WHH$ and $ZHH$ Productions at the HL-LHC}},
  \href{http://arxiv.org/abs/1511.03311}{{\tt 1511.03311}}.

\bibitem{ATL-PHYS-PUB-2014-019}
\emph{{Prospects for measuring Higgs pair production in the channel
  $H(\rightarrow\gamma\gamma)H(\rightarrow b\overline{b}) $ using the ATLAS
  detector at the HL-LHC}},  Tech. Rep. ATL-PHYS-PUB-2014-019, CERN, Geneva,
  Oct, 2014.

\bibitem{ATL-PHYS-PUB-2015-046}
\emph{{Higgs Pair Production in the $H(\rightarrow \tau\tau)H(\rightarrow
  b\bar{b})$ channel at the High-Luminosity LHC}},  Tech. Rep.
  ATL-PHYS-PUB-2015-046, CERN, Geneva, Nov, 2015.

\bibitem{Aad:2015xja}
{\scshape ATLAS} collaboration, G.~Aad et~al., \emph{{Searches for Higgs boson
  pair production in the $hh\to bb\tau\tau, \gamma\gamma WW^*, \gamma\gamma bb,
  bbbb$ channels with the ATLAS detector}},
  \href{http://dx.doi.org/10.1103/PhysRevD.92.092004}{\emph{Phys. Rev.} {\bf
  D92} (2015) 092004}, [\href{http://arxiv.org/abs/1509.04670}{{\tt
  1509.04670}}].

\bibitem{Aad:2015uka}
{\scshape ATLAS} collaboration, G.~Aad et~al., \emph{{Search for Higgs boson
  pair production in the $b\bar{b}b\bar{b}$ final state from pp collisions at
  $\sqrt{s} = 8$ TeVwith the ATLAS detector}},
  \href{http://dx.doi.org/10.1140/epjc/s10052-015-3628-x}{\emph{Eur. Phys. J.}
  {\bf C75} (2015) 412}, [\href{http://arxiv.org/abs/1506.00285}{{\tt
  1506.00285}}].

\bibitem{Khachatryan:2016sey}
{\scshape CMS} collaboration, V.~Khachatryan et~al., \emph{{Search for two
  Higgs bosons in final states containing two photons and two bottom quarks}},
  \href{http://arxiv.org/abs/1603.06896}{{\tt 1603.06896}}.

\bibitem{Plehn:2005nk}
T.~Plehn and M.~Rauch, \emph{{The quartic higgs coupling at hadron colliders}},
  \href{http://dx.doi.org/10.1103/PhysRevD.72.053008}{\emph{Phys. Rev.} {\bf
  D72} (2005) 053008}, [\href{http://arxiv.org/abs/hep-ph/0507321}{{\tt
  hep-ph/0507321}}].

\bibitem{Binoth:2006ym}
T.~Binoth, S.~Karg, N.~Kauer and R.~Ruckl, \emph{{Multi-Higgs boson production
  in the Standard Model and beyond}},
  \href{http://dx.doi.org/10.1103/PhysRevD.74.113008}{\emph{Phys. Rev.} {\bf
  D74} (2006) 113008}, [\href{http://arxiv.org/abs/hep-ph/0608057}{{\tt
  hep-ph/0608057}}].

\bibitem{McCullough:2013rea}
M.~McCullough, \emph{{An Indirect Model-Dependent Probe of the Higgs
  Self-Coupling}}, \href{http://dx.doi.org/10.1103/PhysRevD.90.015001,
  10.1103/PhysRevD.92.039903}{\emph{Phys. Rev.} {\bf D90} (2014) 015001},
  [\href{http://arxiv.org/abs/1312.3322}{{\tt 1312.3322}}].

\bibitem{Goertz:2014qta}
F.~Goertz, A.~Papaefstathiou, L.~L. Yang and J.~Zurita, \emph{{Higgs boson pair
  production in the D=6 extension of the SM}},
  \href{http://dx.doi.org/10.1007/JHEP04(2015)167}{\emph{JHEP} {\bf 04} (2015)
  167}, [\href{http://arxiv.org/abs/1410.3471}{{\tt 1410.3471}}].

\bibitem{Kuhn:2013zoa}
J.~H. K\"uhn, A.~Scharf and P.~Uwer, \emph{{Weak Interactions in Top-Quark Pair
  Production at Hadron Colliders: An Update}},
  \href{http://dx.doi.org/10.1103/PhysRevD.91.014020}{\emph{Phys. Rev.} {\bf
  D91} (2015) 014020}, [\href{http://arxiv.org/abs/1305.5773}{{\tt
  1305.5773}}].

\bibitem{Beneke:2015lwa}
M.~Beneke, A.~Maier, J.~Piclum and T.~Rauh, \emph{{Higgs effects in top
  anti-top production near threshold in $e^+e^-$ annihilation}},
  \href{http://dx.doi.org/10.1016/j.nuclphysb.2015.07.034}{\emph{Nucl. Phys.}
  {\bf B899} (2015) 180--193}, [\href{http://arxiv.org/abs/1506.06865}{{\tt
  1506.06865}}].

\bibitem{Gorbahn:2016uoy}
M.~Gorbahn and U.~Haisch, \emph{{Indirect probes of the trilinear Higgs
  coupling: $gg \to h$ and $h \to \gamma \gamma$}},
  \href{http://arxiv.org/abs/1607.03773}{{\tt 1607.03773}}.

\bibitem{Hahn:2000kx}
T.~Hahn, \emph{{Generating Feynman diagrams and amplitudes with FeynArts 3}},
  \href{http://dx.doi.org/10.1016/S0010-4655(01)00290-9}{\emph{Comput. Phys.
  Commun.} {\bf 140} (2001) 418--431},
  [\href{http://arxiv.org/abs/hep-ph/0012260}{{\tt hep-ph/0012260}}].

\bibitem{Hahn:1998yk}
T.~Hahn and M.~Perez-Victoria, \emph{{Automatized one loop calculations in
  four-dimensions and D-dimensions}},
  \href{http://dx.doi.org/10.1016/S0010-4655(98)00173-8}{\emph{Comput. Phys.
  Commun.} {\bf 118} (1999) 153--165},
  [\href{http://arxiv.org/abs/hep-ph/9807565}{{\tt hep-ph/9807565}}].

\bibitem{Mertig:1990an}
R.~Mertig, M.~Bohm and A.~Denner, \emph{{FEYN CALC: Computer algebraic
  calculation of Feynman amplitudes}},
  \href{http://dx.doi.org/10.1016/0010-4655(91)90130-D}{\emph{Comput. Phys.
  Commun.} {\bf 64} (1991) 345--359}.

\bibitem{Shtabovenko:2016sxi}
V.~Shtabovenko, R.~Mertig and F.~Orellana, \emph{{New Developments in FeynCalc
  9.0}},  \href{http://arxiv.org/abs/1601.01167}{{\tt 1601.01167}}.

\bibitem{Aglietti:2004nj}
U.~Aglietti, R.~Bonciani, G.~Degrassi and A.~Vicini, \emph{{Two loop light
  fermion contribution to Higgs production and decays}},
  \href{http://dx.doi.org/10.1016/j.physletb.2004.06.063}{\emph{Phys. Lett.}
  {\bf B595} (2004) 432--441}, [\href{http://arxiv.org/abs/hep-ph/0404071}{{\tt
  hep-ph/0404071}}].

\bibitem{Degrassi:2004mx}
G.~Degrassi and F.~Maltoni, \emph{{Two-loop electroweak corrections to Higgs
  production at hadron colliders}},
  \href{http://dx.doi.org/10.1016/j.physletb.2004.09.008}{\emph{Phys. Lett.}
  {\bf B600} (2004) 255--260}, [\href{http://arxiv.org/abs/hep-ph/0407249}{{\tt
  hep-ph/0407249}}].

\bibitem{Actis:2008ug}
S.~Actis, G.~Passarino, C.~Sturm and S.~Uccirati, \emph{{NLO Electroweak
  Corrections to Higgs Boson Production at Hadron Colliders}},
  \href{http://dx.doi.org/10.1016/j.physletb.2008.10.018}{\emph{Phys. Lett.}
  {\bf B670} (2008) 12--17}, [\href{http://arxiv.org/abs/0809.1301}{{\tt
  0809.1301}}].

\bibitem{Degrassi:2010eu}
G.~Degrassi and P.~Slavich, \emph{{NLO QCD bottom corrections to Higgs boson
  production in the MSSM}},
  \href{http://dx.doi.org/10.1007/JHEP11(2010)044}{\emph{JHEP} {\bf 11} (2010)
  044}, [\href{http://arxiv.org/abs/1007.3465}{{\tt 1007.3465}}].

\bibitem{Degrassi:2005mc}
G.~Degrassi and F.~Maltoni, \emph{{Two-loop electroweak corrections to the
  Higgs-boson decay $H \to \gamma \gamma$}},
  \href{http://dx.doi.org/10.1016/j.nuclphysb.2005.06.027}{\emph{Nucl. Phys.}
  {\bf B724} (2005) 183--196}, [\href{http://arxiv.org/abs/hep-ph/0504137}{{\tt
  hep-ph/0504137}}].

\bibitem{Actis:2008ts}
S.~Actis, G.~Passarino, C.~Sturm and S.~Uccirati, \emph{{NNLO Computational
  Techniques: The Cases $H \to \gamma \gamma$ and $H \to g g$}},
  \href{http://dx.doi.org/10.1016/j.nuclphysb.2008.11.024}{\emph{Nucl. Phys.}
  {\bf B811} (2009) 182--273}, [\href{http://arxiv.org/abs/0809.3667}{{\tt
  0809.3667}}].

\bibitem{MelladoGarcia:2150771}
B.~Mellado~Garcia, P.~Musella, M.~Grazzini and R.~Harlander, \emph{{CERN Report
  4: Part I Standard Model Predictions}}, .

\bibitem{Butterworth:2015oua}
J.~Butterworth et~al., \emph{{PDF4LHC recommendations for LHC Run II}},
  \href{http://dx.doi.org/10.1088/0954-3899/43/2/023001}{\emph{J. Phys.} {\bf
  G43} (2016) 023001}, [\href{http://arxiv.org/abs/1510.03865}{{\tt
  1510.03865}}].

\bibitem{Dulat:2015mca}
S.~Dulat, T.-J. Hou, J.~Gao, M.~Guzzi, J.~Huston, P.~Nadolsky et~al.,
  \emph{{New parton distribution functions from a global analysis of quantum
  chromodynamics}},
  \href{http://dx.doi.org/10.1103/PhysRevD.93.033006}{\emph{Phys. Rev.} {\bf
  D93} (2016) 033006}, [\href{http://arxiv.org/abs/1506.07443}{{\tt
  1506.07443}}].

\bibitem{Harland-Lang:2014zoa}
L.~A. Harland-Lang, A.~D. Martin, P.~Motylinski and R.~S. Thorne, \emph{{Parton
  distributions in the LHC era: MMHT 2014 PDFs}},
  \href{http://dx.doi.org/10.1140/epjc/s10052-015-3397-6}{\emph{Eur. Phys. J.}
  {\bf C75} (2015) 204}, [\href{http://arxiv.org/abs/1412.3989}{{\tt
  1412.3989}}].

\bibitem{Ball:2014uwa}
{\scshape NNPDF} collaboration, R.~D. Ball et~al., \emph{{Parton distributions
  for the LHC Run II}},
  \href{http://dx.doi.org/10.1007/JHEP04(2015)040}{\emph{JHEP} {\bf 04} (2015)
  040}, [\href{http://arxiv.org/abs/1410.8849}{{\tt 1410.8849}}].

\bibitem{Grojean:2004xa}
C.~Grojean, G.~Servant and J.~D. Wells, \emph{{First-order electroweak phase
  transition in the standard model with a low cutoff}},
  \href{http://dx.doi.org/10.1103/PhysRevD.71.036001}{\emph{Phys. Rev.} {\bf
  D71} (2005) 036001}, [\href{http://arxiv.org/abs/hep-ph/0407019}{{\tt
  hep-ph/0407019}}].

\end{thebibliography}\endgroup

\end{document}